# Cavitation Dynamics in Venturi-Type Microchannels: Experimental Observations and Numerical Modeling

Md Naim Hossain[a], Anish Pal[a], Sayan Karmakar[a], Alexander D. Snyder[b], Michael T. Barako[b], Clayton A. Pullins[b], Constantine M. Megaridis[a,*]

[a] *Mechanical and Industrial Engineering, University of Illinois Chicago, Chicago, IL, United States*

[b] *Northrop Grumman Corporation, Baltimore, MD, United States*

* cmm@uic.edu

## ABSTRACT

Cavitation in fluids can severely hinder the efficiency of the associated flows. This undesired phenomenon is strongly influenced by local flow conditions, flow orientation, proximity to boundaries and liquid/gas properties at saturation. When liquid flow is severely constricted and cavitation occurs, diverging microchannels can suppress vapor formation through effective pressure recovery; however, the behavior of such microchannels under different geometric and operating conditions remains unclear. This work combines experimental data and computational modeling to elucidate the intricate flow dynamics of cavitation-induced phase change of refrigerant (R134a) flow in a Venturi-type microchannel. Full-field numerical simulations are carried out using the interPhaseChangeFoam solver in OpenFOAM®, with the model coefficients being validated against experimental data obtained with an identical system. The effects of flow rate, channel opening angle, and inlet cross-sectional geometry are explored with a focus on pressure drop, velocity distribution, and cavitation characteristics, including vapor fraction. The results highlight the roles of flow velocity, orifice-to-channel width ratio, orifice size, and channel divergence angle on overall pressure drop. A channel opening angle of 12° minimizes pressure drop, in contrast with conventional single-phase Venturi geometries where the optimal opening angle typically lies between 5° and 7°. The study underscores that the cavitation number, including its sign, serves as a critical quantitative indicator of cavitation severity—where increasingly negative values correspond to intensified vapor generation in microscale constrictions. By identifying the relationships between operating conditions and cavitation phenomena, this work provides a framework for understanding and optimizing microchannel designs under conditions prone to cavitation.

# NOMENCLATURE

| | | | |
|---|---|---|---|
| $C_C$ | Cavitation accommodation coefficient for condensation | $V_{in}$ | Mean pipe-inlet velocity |
| $C_V$ | Cavitation accommodation coefficient for evaporation | $\dot{V}_C$ | Volume conversion rate for condensation |
| $C_\mu$ | Empirical constant | $\dot{V}_V$ | Volume conversion rate for vaporization |
| $CD_{k\omega}$ | Positive portion of the cross-diffusion term | $W_n$ | Throat width |
| D | Inlet pipe diameter | $x_i$ | Cartesian coordinate in i[th] direction |
| d | Orifice width | **Greek** | |
| $d^*$ | Nearest wall distance | $\alpha_l$ | Liquid volume fraction |
| $d_{out}$ | Channel outlet width | $\alpha_v$ | Vapor volume fraction |
| $F_{1,2}$ | Blending functions (Eqs. 11, 12) | $\alpha_{nuc}$ | Cavitation initiation nucleation factor |
| $g_i$ | Gravitational acceleration along *i*-direction | β | d/D ratio |
| H | Height of channel | $\delta_{ij}$ | Kronecker-delta tensor |
| k | Turbulent kinetic energy | θ | Channel divergence (opening) angle |
| l | Characteristic length | μ | Dynamic viscosity |
| $\dot{m}_C$ | Mass conversion rate due to condensation | $\mu_t$ | Eddy viscosity |
| $\dot{m}_V$ | Mass conversion rate due to vaporization | ρ | Density of liquid-vapor mixture |
| $n_b$ | Number density of nuclei/bubble per unit volume | $\rho_l$ | Liquid phase density |
| P | Turbulent production term (Eq. 7) | $\rho_m$ | Mixture density |
| p | Pressure | $\rho_v$ | Vapor phase density |
| $p_{out}$ | Outlet pressure | σ | Cavitation Number |
| $p_{sat}$ | Saturation pressure | $\sigma_k$ | Constant for turbulent diffusion of *k* |
| Q | Volume flow rate | $\sigma_\omega$ | Constant for turbulent diffusion of *ω* |
| $R_B$ | Vapor bubble radius | $\tau_{ij}$ | Stress tensor |
| $S_{ij}$ | Strain rate (Eq. 4) | ϕ | Thermophysical property |
| t | Time | ω | Turbulent energy dissipation rate |
| $t_c$ | Channel thickness | $\Omega_{ij}$ | Vorticity tensor |
| $u_i$ | Velocity component in the i[th] direction | | |

# 1. Introduction

Cavitation is manifested by the instantaneous formation and rapid collapse of vapor bubbles in a liquid driven under high flow rates through physical constrictions, where local pressures can drop below the saturated vapor pressure of the fluid. This dynamic process has been studied extensively due to its critical implications in areas such as propulsion systems, thermal management devices, and microfluidics [1–6]. The cavitation number, a dimensionless measure of pressure-to-inertial force ratio, dictates the susceptibility and regime of cavitation, with lower cavitation numbers associated with higher cavitation risk. The phenomenon introduces unique complexities driven by the interplay of surface tension, viscous forces, and thermal gradients, which can modify the flow behavior [7,8]. This poses operational challenges, ranging from increased flow resistance and reduced heat exchange efficiency to unstable operation and catastrophic hardware damage caused by the formation and implosion of vapor bubbles [9–12]. Diverging microchannels offer a mitigation strategy to suppress cavitation via widening cross section along the flow path, which modulates pressure drop, and thus, reduces cavitation risk. Despite the volume of prior works on cavitation, studies focusing on diverging microchannels remain limited.

To better understand and control cavitation in microchannels, researchers have employed a combination of numerical simulations and experimental investigations. High-fidelity computational methods—such as large-eddy simulations (LES) and compressible RANS solvers with preconditioning—have been developed to simulate turbulent cavitating flows and validate homogeneous mixture models across representative throttle microchannel and Venturi-type geometries [13–15]. These simulations provided valuable insights into the interaction between cavitation structures and coherent flow motions, the evolution of turbulence under phase change, and the utility of different equations of state in capturing vapor–liquid dynamics. Nearly all prior work addressed cavitation in a manner that was highly specific to the system under study.

Complementing the modeling efforts, experimental studies have explored the transient evolution of cavitating flows under various geometries. Investigations in converging-diverging nozzles and rectangular microchannels have revealed distinct cavitation regimes, such as quasi-isothermal and thermo-sensitive transitions, while highlighting the sensitivity of cavitation intensity and vapor distribution to taper angles, inlet cross-sectional area, and the presence of inlet prechambers [8,16–18]. In particular, microfluidic test platforms integrated with visualization

3
Approved for Public Release: NG25-2102. ©2026 Northrop Grumman Systems Corporation

techniques have enabled real-time observation of cavitation bubble evolution. These studies demonstrated that lower cavitation numbers intensify vapor formation, increasing bubble distribution and jet length, whereas double flow restriction structures suppress cavitation more effectively than single restrictions [9,18]. Nonetheless, experimental instrumentation is inherently limited in keeping up with the rapid dynamic phenomena in cavitating flows, thus making their implementation in such flows intractable. Beyond geometry, boundary surface properties have been shown to add another layer of complexity to microscale cavitation. Surface roughness and wettability critically influence local flow fields, hydraulic resistance, and nucleation dynamics. For instance, rough surfaces tend to enhance cavitation by promoting vapor nucleation, as surface asperities and crevices act as favorable sites for bubble inception, while surface treatments can localize and modulate bubble growth [19,20]. These findings underscore the importance of surface engineering in microscale cavitation control. In addition to structural and surface parameters, operating conditions, particularly flow velocity and cavitation number, play a defining role in cavitation behavior. Variations in flow rate (and likewise, velocity) and cavitation number have been observed to affect vapor cloud length, collapse frequency, and turbulence modulation, directly impacting flow stability [13,19,21].

Synthesizing these insights, recent works have proposed design guidelines for cavitation mitigation in microchannels with an emphasis on optimizing geometric transitions, surface properties, and flow conditions [22,23]. Among various configurations, diverging microchannels integrated with prechambers have attracted significant attention for their effectiveness in stabilizing two-phase flows and controlling phase-change dynamics [17,24]. Together, these studies offer a comprehensive foundation for the rational design of cavitation-resilient microfluidic systems aimed at improving performance and reliability in engineering applications.

This work builds upon prior foundational research on cavitation by systematically investigating cavitation phenomena in diverging microchannel integrated with pre-chambers separated by a severe constriction (orifice). The rigorous experimentation concentrates on cavitation dynamics of *turbulent* flows through diverging microchannels designed for electronics cooling applications, especially those involving vertically integrated, heterogeneous chip stacks [25]. The specific focus is on understanding the pressure drop characteristics and the onset of cavitation under varying flow conditions and geometric configurations of the microchannel. By systematically analyzing these factors, the study provides new insights into optimizing microchannel designs to minimize cavitation risks and ensure effective flow management in advanced electronic systems [26]. Numerical simulations performed on high-performance parallel computing resources use OpenFOAM® to analyze the complex interactions between geometric features and flow conditions, with simulation results benchmarked against the experimental data. The Schnerr–Sauer model [22,27,28], widely recognized for transient cavitation modeling, is coupled with the interPhaseChangeFOAM solver in OpenFOAM® to develop a numerical model of the flow domain from the pre-chamber entry to the exit of the diverging microchannel. The research identifies key influencing parameters for entry-to-outlet pressure drop, vapor distribution, and flow dynamics within the channel. These parameters include the geometry of the diverging section, particularly the opening angle and the ratio of the orifice width to the entry-chamber width. The results provide an example for optimizing the design of diverging microchannels towards maximum flow rate with minimal cavitation risk. The combined experimental and numerical approaches provide a generalizable framework for modeling, predicting and interpreting cavitation in microchannels, thus advancing the design and optimization of such systems [29–31] for applications ranging from biomedicine, materials engineering and thermal management to chemical processing.

## 2. System Definition

### 2.1 Channel Geometry

The flow system consists of a cylindrical supply tube attached vertically to a low-profile prechamber connected to a laterally diverging microchannel through a small orifice as illustrated in **Fig. 1**. The prechamber has a diameter $D$ (matching the pre-chamber width) and serves as the initial fluid facilitator. Fluid enters from the top and is directed toward the orifice (width $d$) which acts as a flow constriction between the pre-chamber and the microchannel. The microchannel gradually expands from the orifice to the outlet, where it reaches a final width of $d_{out}$. The channel maintains a constant height $t_c$, which defines the depth of the channel. Fluid flow follows the direction indicated by the blue arrow in **Fig. 1**, moving from the pre-chamber through the orifice, along the microchannel, and finally exiting at the wide outlet. This configuration is well suited for exploring flow behavior and pressure distribution in constricted microfluidic systems and has the advantage of post-orifice pressure recovery, which suppresses cavitation in the region beyond the narrow channel entry.

The parametric values and geometric dimensions considered in the present study are summarized in **Table 1**. These values have been carefully selected to cover a wide range of flow conditions and microchannel geometries, particularly for a cooling path within a multi-die vertically integrated heterogeneous chip stack. A systematic analysis of the selected parameter ranges was conducted to evaluate their influence on flow behavior, pressure drop, and cavitation characteristics, providing a comprehensive understanding of microchannel performance under various design and operating conditions.

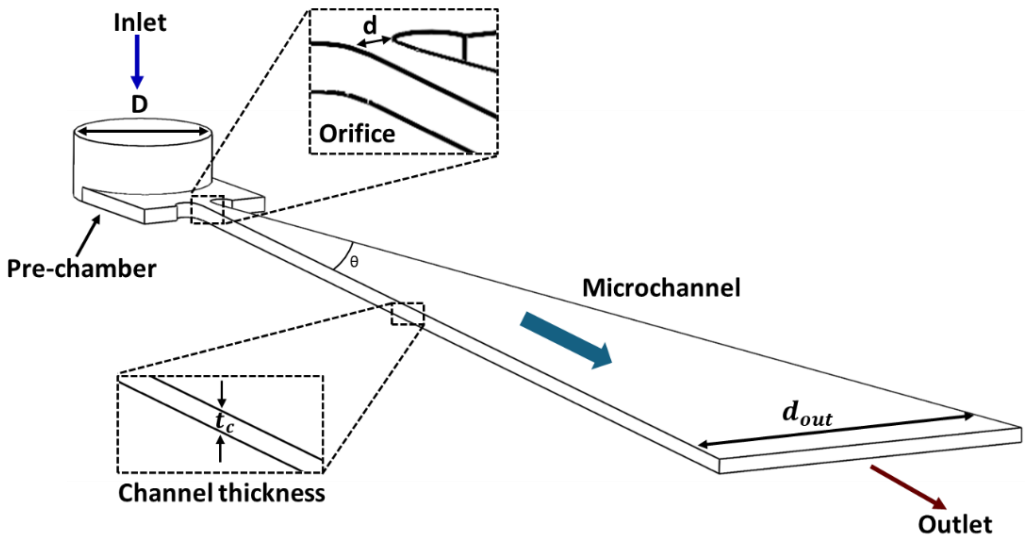

**Figure 1.** Geometry of the microchannel fluid domain with the prechamber attached to the vertical entry tube. The schematic shows the boundaries of the fluid flow. The vertical entry tube is shown shortened for illustrative clarity; its actual length extends beyond the portion depicted.

**Table 1.** Geometric and flow parameter ranges considered in this work

| Parameter | Symbol | Range / Value | Units |
|---|---|---|---|
| Prechamber width | $D$ | 2.794 | mm |
| Orifice width | $d$ | $325 - 505$ | μm |
| Channel height (Vertical thickness) | $t_c$ | 300 | μm |
| Channel outlet width | $d_{out}$ | $1.48 - 8.98$ | mm |
| Mass flow rate | $\dot{m}$ | $1.7 - 6.5$ | g/s |
| Inlet Pressure | $p$ | $13.5 \times 10^5$ | Pa |

## 2.2 Mathematical Model

The governing equations for a system involving two immiscible fluids undergoing phase change apply here. The mass continuity equation has the form [32,33]

$$\frac{\partial \rho}{\partial t} + \frac{\partial}{\partial x_j}(\rho u_j) = 0 \qquad (1)$$

while the multiphase momentum equation [34,35]

$$\rho \frac{\partial u_i}{\partial t} + \frac{\partial}{\partial x_j}(\rho u_i u_j) = -\frac{\partial p}{\partial x_j} + \frac{\partial \tau_{ij}}{\partial x_i} + \rho g_i + F_{B_i} \qquad (2)$$

is Reynolds-averaged [36–38] in conjunction with the $k$-$\omega$ turbulence model in the shear stress formulation [39]. $F_{B_i}$ denotes the body force term in the momentum equation and includes the effect of surface tension through the methodology as proposed by Brackbill et al. [33].

The closure equation for the momentum equation has the form

$$\tau_{ij} = \mu_t \left( 2 S_{ij} - \frac{2}{3} \frac{\partial u_k}{\partial x_k} \delta_{ij} \right) - \frac{2}{3} \rho k \delta_{ij} \qquad (3)$$

where the strain rate is

$$S_{ij} = \frac{1}{2}\left( \frac{\partial u_i}{\partial x_j} + \frac{\partial u_j}{\partial x_i} \right) \qquad (4)$$

The transport equation for the turbulent kinetic energy $k$ is

$$\frac{\partial (\rho k)}{\partial t} + \frac{\partial (\rho u_j k)}{\partial x_j} = P - \beta^* \rho \omega k + \frac{\partial}{\partial x_j}\left[ (\mu + \sigma_k \mu_t) \frac{\partial k}{\partial x_j} \right] \qquad (5)$$

while the specific dissipation rate $\omega$ obeys this equation

$$\frac{\partial (\rho \omega)}{\partial t} + \frac{\partial (\rho u_j \omega)}{\partial x_j} = P - \beta' \rho \omega^2 + \frac{\partial}{\partial x_j}\left[ (\mu + \sigma_\omega \mu_t) \frac{\partial \omega}{\partial x_j} \right] \qquad (6)$$

The turbulent production term is

$$P = \tau_{ij} \frac{\partial u_i}{\partial x_j} \qquad (7)$$

and the eddy viscosity limiter is

$$\mu_t = \frac{\rho a_1 k}{\max(a_1 \omega, \Omega_{ij} F_2)} \tag{8}$$

where

$$\Omega_{ij} = \frac{1}{2}\left(\frac{\partial u_i}{\partial x_j} - \frac{\partial u_j}{\partial x_i}\right) \tag{9}$$

The model constants are weighted per

$$(\sigma_k, \sigma_\omega, \beta')^T = F_1 (\sigma_k, \sigma_\omega, \beta')_1^T + (1 - F_1)(\sigma_k, \sigma_\omega, \beta')_2^T \tag{10}$$

where the blending functions are

$$F_1 = \tanh\left\{\left[\min\left[\max\left(\frac{\sqrt{k}}{\beta^* \omega d}, \frac{500\nu}{d^{*2}\omega}\right), \frac{4\rho\sigma_{\omega 2} k}{CD_{k\omega} d^{*2}}\right]\right]^4\right\} \tag{11}$$

and

$$F_2 = \tanh\left\{\left[\max\left(2\frac{\sqrt{k}}{\beta^* \omega d^*}, \frac{500\nu}{d^{*2}\omega}\right)\right]^2\right\} \tag{12}$$

with

$$CD_{k\omega} = \max\left(\frac{2\rho}{\sigma_{\omega 2}\omega}\frac{\partial k}{\partial x_j}\frac{\partial \omega}{\partial x_j}, 10^{-20}\right) \tag{13}$$

The low-end capping of $CD_{k\omega}$ ensures that the last fractional term in Eq. 11 does not overshoot [39].

The above expressions include the following constants obtained from [39]

$$\begin{aligned}
&\sigma_{k1} = 0.85, \quad \sigma_{\omega 1} = 0.65, \quad \beta'_1 = 0.075 \\
&\sigma_{k2} = 1.00, \quad \sigma_{\omega 2} = 0.856, \quad \beta'_2 = 0.0828 \\
&\beta^* = 0.09, \quad a_1 = 0.31
\end{aligned} \tag{14}$$

For multiphase flow, the continuity equation is not implemented directly. Instead, the volume of fluid model (VOF) is implemented in conjunction with the continuity equation for interface tracking. The modified continuity equation for interface tracking determines the liquid volume fraction on the mass transfer rates between the liquid (*l*) and vapor (*v*) phases

$$\frac{\partial(\rho_l \alpha_l)}{\partial t} + \frac{\partial(\rho \alpha_l u_j)}{\partial x_j} = \dot{m}_v + \dot{m}_c \tag{15}$$

where $\alpha_l + \alpha_v = 1$. The mass transfer rates $\dot{m}_v$ (evaporation) and $\dot{m}_c$ (condensation) in Eq. (15) depend on the chosen phase change model. Equation (15) is further modified to obtain

$$\frac{\partial \alpha_l}{\partial t} + \frac{\partial(\alpha_l u_j)}{\partial x_j} - \alpha_l \frac{\partial u_j}{\partial x_j} = \alpha_l (\dot{V}_V - \dot{V}_C) + \dot{V}_C \tag{16}$$

where $\dot{V}_C$ and $\dot{V}_V$ are calculated by

$$\dot{V}_C = \left(\frac{1}{\rho_l} - \alpha_l \left(\frac{1}{\rho_l} - \frac{1}{\rho_v}\right)\right) \dot{m}_C \tag{17}$$

$$\dot{V}_V = \left(\frac{1}{\rho_l} - \alpha_l \left(\frac{1}{\rho_l} - \frac{1}{\rho_v}\right)\right) \dot{m}_V \tag{18}$$

In the above equations, the variables and all other subscripted parameters represent the mean values at the center of each computational cell. Any thermophysical property of the mixture, denoted as $\phi$, is determined from the respective properties of the liquid and vapor phases, weighted by their respective volume fractions, i.e.

$$\phi = \alpha_l \phi_l + \alpha_v \phi_v = \alpha_l \phi_l + (1 - \alpha_l) \phi_l \tag{19}$$

The Schnerr-Sauer cavitation model, which was originally developed for simulating transient cavitation and utilizes the simplified Rayleigh-Plesset bubble dynamics equations, is implemented for mass-transfer modeling [40]

$$\dot{m}_V = C_V (1 + \alpha_{\text{nuc}} - \alpha_l) \alpha_l \frac{3 \rho_l \rho_v}{\rho R_B} \sqrt{\frac{2}{3 \rho_l} \frac{\min|[p - p_{\text{sat}}, 0]|}{|p - p_{\text{sat}} + 0.01 p_{\text{sat}}|}} \tag{20}$$

$$\dot{m}_C = C_C \alpha_l \frac{3 \rho_l \rho_v}{\rho R_B} \sqrt{\frac{2}{3 \rho_l} \frac{\max|[p - p_{\text{sat}}, 0]|}{|p - p_{\text{sat}} + 0.01 p_{\text{sat}}|}} \tag{21}$$

Apart from the empirical cavitation accommodation coefficients, $C_V$ and $C_C$, the above expressions involve a nucleation site volume fraction $\alpha_{\text{nuc}}$, number density of nuclei per unit volume $n_b$ and the bubble radius $R_B$. The respective expressions for $\alpha_{\text{nuc}}$ and $R_B$ are [41]

$$\alpha_{nuc} = \frac{\frac{\pi d_{\text{nuc}}^3 n_b}{6}}{1 + \frac{\pi d_{\text{nuc}}^3 n_b}{6}} \tag{22}$$

$$R_B = \left(\frac{3}{4\pi n_b} \frac{1+\alpha_{nuc}-\alpha_l}{\alpha_l}\right)^{1\backslash 3} \tag{23}$$

## 2.3 Initial and Boundary Conditions

The no-slip boundary condition was imposed on all solid walls of the microchannel. The liquid refrigerant filling the channel was assumed to be initially still until ($t$ = 0) a non-zero velocity was imposed at the entry flow (fully developed). The outlet was assigned a pressure outlet boundary condition, allowing for natural flow adjustment and ensuring realistic exit conditions. There was no heat input to the present system.

To model cavitation, the Schnerr-Sauer cavitation model was implemented, with the cavitation accommodation coefficient tuned using the experimental data (see Section 3.2). This coefficient is calibrated to ensure accurate prediction of phase-change dynamics and pressure-drop behavior.

## 2.4 Experimental Methods

Hydraulic flow characterization was conducted for diverging channel geometries with orifice widths of 325 μm, 355 μm, and 505 μm in addition to the following fixed parameters: 300 μm height, 30.4 mm orifice-to-outlet distance, and 7.61 mm outlet width. The fixed outlet width and channel length, along with the variable entry width translate to slightly different divergence angles for these three channel geometries. The test vehicle was constructed using a laser-cut copper shim to form the channel geometry, sealed between a milled 6061 aluminum base and polyetherimide upper (Ultem 1000®, SABIC) with Buna-N seals. **Figure 2** depicts the test vehicle, starting with the copper shim (a) and then the top (b) and side (c) views of the full assembly.

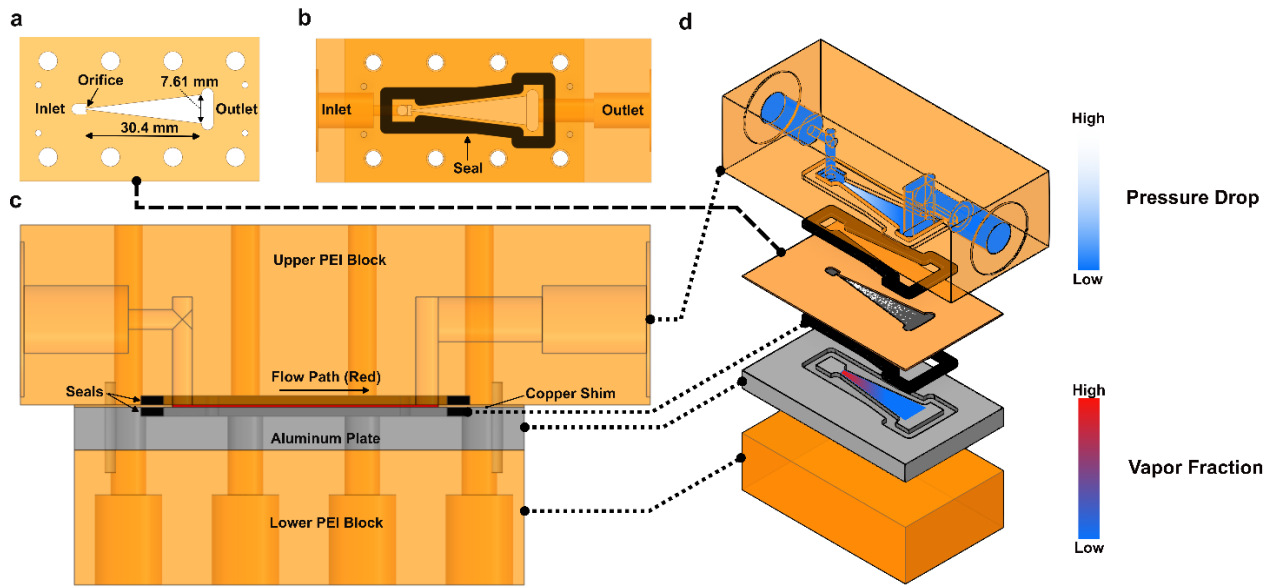

**Figure 2.** Hydraulic test vehicle assembly. (a) Top view of the copper shim used to set the channel geometry, with points and fixed dimensions of interest shown. (b) Top view of the test vehicle assembly showing the position of the flow path relative to the seal surfaces. (c) Midline cross section of the test vehicle assembly with components labeled. (d) 3-D depiction of the flow path (top) and exploded view (middle and bottom) of the present microchannel experimental assembly and its layered components. At top, the refrigerant (in blue) enters horizontally from the left and exits from the right.

Refrigerant flow tests were conducted on a high-pressure loop utilizing a Coriolis flow meter (Micromotion CMF-010, Emerson®) for volumetric flow rate monitoring in conjunction with a cavity style gear pump (74445-21, Cole-Parmer®, Inc.) connected to a process controller (CNi3253, DWYEROMEGA™), data acquisition (DAQ) unit (NI-USB 6525™), and 5V power supply (PS-S6024, Altech®) to form a Proportional–Integral–Derivative (PID) loop for precise flow rate control. The test vehicle's pressure drop was monitored using pressure transducers placed near the inlet and outlet (PX309-200A-10V, DWYEROMEGA™) and a differential pressure transducer plumbed in parallel to the flow path (409-100DWU10V, DWYEROMEGA™). All transducers had a stated accuracy of ± 0.25%. Fluid temperature was monitored using thermocouples plumbed at the test piece inlet and outlet (TMQSS-125V-6, DWYEROMEGA™), with a brazed plate double-wall heat exchanger (BP-415-034, Bell & Gossett®) and 50-50 ethylene glycol and water (EGW) mix cooling cart (59660, Electro-Impulse Inc.) used to hold the inlet fluid temperatures at ≈ 50°C and ensure full condensation of the fluid after passing through the test piece. A vertical head distance of ≈ 0.75 m between the pump and the test piece inlet was leveraged to ensure vapor-free fluid delivery prior to the channel orifice.

Flow resistance curves were constructed for each orifice by sweeping volumetric flow rates of 0.025 gpm, 0.035 gpm, and (where possible) 0.046 gpm (1.7, 2.4, and 3.2 g/s), while holding the inlet temperature fixed at ≈ 50 °C. Five pressure drop readings were collected at 1 Hz for each flow rate point using a DAQ module (NI-USB-6001, NI$^{TM}$) and a custom MATLAB® script for data monitoring and acquisition.

## 3. Results and Discussion

### 3.1 Model Spatial Grid and Time-step Independence Study

A grid and time-step independence study ensured that the numerical results are free from discretization errors and adequately capture the physics of the problem without excessive computational cost. A 3D structured, hexahedral-dominant mesh was used for the discretization of the computational domain. Five different mesh resolutions were considered: a very coarse mesh with approximately $1.7 \times 10^6$ cells, a coarse mesh with around $5.7 \times 10^6$ cells, a medium mesh with $7.6 \times 10^6$ cells, a medium-fine mesh with approximately $12 \times 10^6$ cells, and a fine mesh with about $20 \times 10^6$ cells. Mesh refinement was applied near the orifice, outlets, and other critical boundaries, ensuring that key flow characteristics are captured accurately. Most importantly, sufficient refinement was applied near the orifice, where flow acceleration and phase interactions are most significant. A smooth and gradual transition was implemented from the highly refined region near the orifice to the coarser mid-channel mesh. The medium mesh (approximately $7.6 \times 10^6$ cells) and the corresponding refinement zones are illustrated in **Fig. 3**. The final mesh resolution was selected to balance computational cost and accuracy. Grid-mesh independence was assessed while keeping the time step fixed, whereas time-step independence was analyzed by refining $\Delta t$ for the grid designated from the grid-independence study.

The grid and time-step independence runs were conducted for a microchannel base-case configuration with an opening size of 355 μm (width) by 300 μm (height). The simulations were performed at different grid resolutions using a fixed timestep of $10^{-6}$ s. The pressure drop along the channel was calculated for each case and constituted the criterion for grid independence. Similarly, for time-step independence, simulations were conducted using a grid size of 20 μm (medium mesh, ~7.6×10⁶ cells) with refinement near the orifice, and the corresponding pressure drop values were compared.

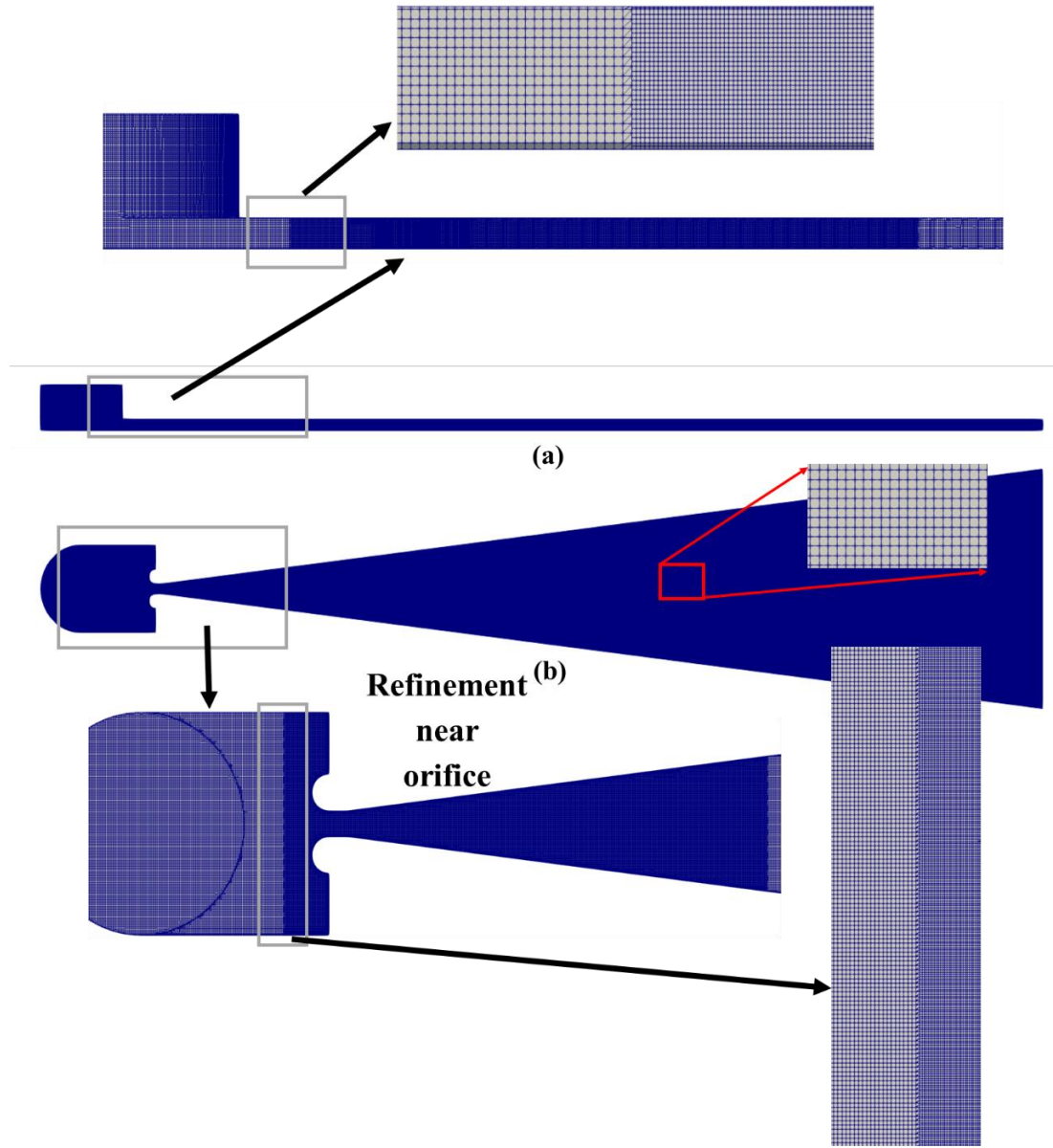

**Figure 3.** Side view (a) and top view (b) of the computational mesh (medium resolution) showing discretization refinement near the orifice. The insets show magnifications of the bracketed domains.

### 3.1.1 Computational Grid Independence

Five different grid resolutions were considered with base spacings of 25 μm, 20 μm, 10 μm and 8 μm, with some also featuring select local refinement. These five grids had approximately $1.7 \times 10^6$ (very coarse), $5.7×10^6$ (coarse), $7.6×10^6$ (medium), $12×10^6$ (medium-fine), and $20×10^6$ (fine) cells, respectively. For the first two grids (25 μm and 20 μm base spacings), no local refinement was made for the entire domain. In the third grid (20 μm base with refinement), mesh refinement was applied near the orifice region, reducing the local cell size to approximately 3 μm. The final two meshes (10 μm and 8 μm) also incorporated refinement in critical regions. The medium mesh (7.6 million cells) yielded results closely matching those of the finer meshes. The pressure drop from the tube entry to the channel's outlet was computed for each case, and the results are presented in **Fig. 4**. As the grid was refined, a notable reduction in deviation from the coarser grid was observed. Beyond 7.6 million cells, any further refinement led to minimal differences in the calculated pressure drop. Consequently, the $7.6×10^6$ cell mesh was selected for all subsequent calculations.

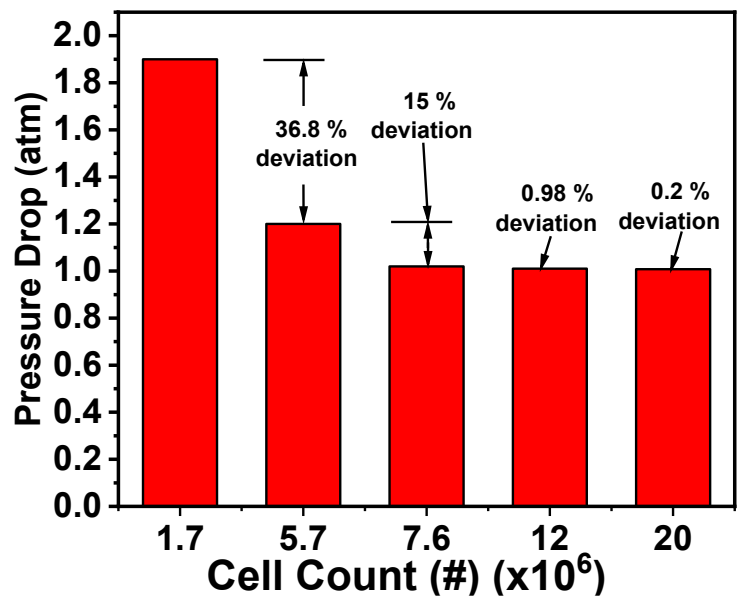

**Figure 4.** Pressure drop for different computational cell counts. Refinement beyond the $7.6×10^6$ mesh provided no further benefit, and thus, this mesh was used for all subsequent runs.

### 3.1.2 Time-Step Independence Study

Following the selection of the spatial grid, simulations were performed with the following time steps: $10^{-4}$ s, $5×10^{-5}$ s, $10^{-5}$ s, $5×10^{-6}$ s, $10^{-6}$ s, and $5×10^{-7}$ s. The pressure drop results and percentage deviations for each timestep are summarized in **Fig. 5**. Large variations in pressure drop are observed at coarser timesteps, whereas at $10^{-6}$ s the deviation reduces to 0.99%, signifying negligible change with further refinement. Beyond this, additional timestep reduction produces no meaningful improvement, and thus $10^{-6}$ s was chosen as the optimal balance between accuracy and computational efficiency.

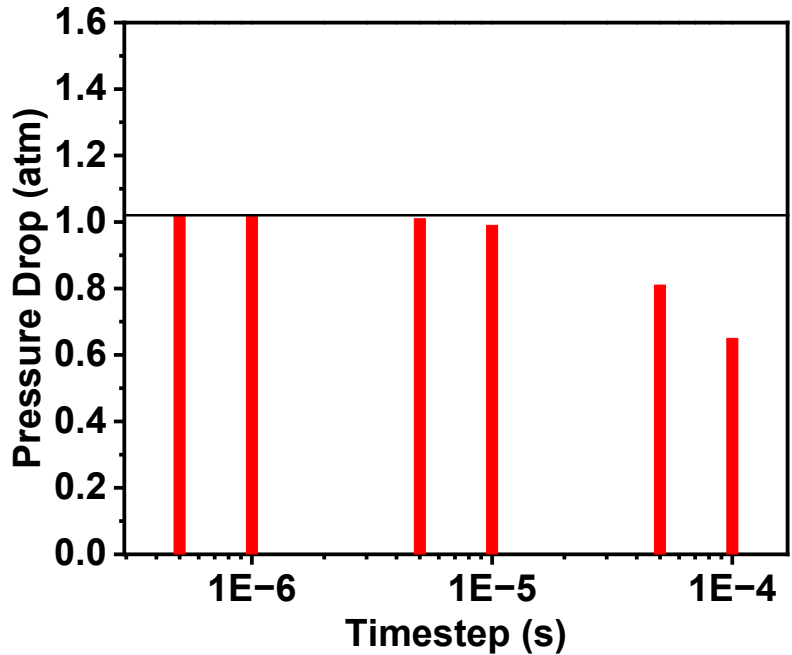

**Figure 5.** Pressure drop along the channel (tube entry to channel exit) for the six different time steps used in the calculations conducted with $7.6 \times 10^6$ computational cells. The horizontal black line indicates $\Delta p$ for the optimal time step of $10^{-6}$ s.

### 3.2 Model Validation

The cavitation accommodation coefficients ($C_V$ and $C_C$) were tuned against the experimental data obtained under the same geometric and flow operating conditions. This tuning was performed for a microchannel with an orifice opening of 325 μm × 300 μm at a mass flow rate of 3.2 g/s. Different values of the accommodation coefficient (0.01 to 1) were tested, and the pressure drop obtained from simulations was compared with the experimental results. As shown in **Fig. 6(a),** the best match was achieved at $C_V = C_c = 0.04$, which was then fixed for subsequent validation cases. This initial tuning at a high mass flow rate ensured that the model accurately captured the cavitation dynamics under flow conditions where cavitation risk is highest (Furthermore, it was found that maintaining a non-zero accommodation coefficient up to a length of ten orifice widths is sufficient to obtain the desired results; additional details are provided in Appendix A1.).

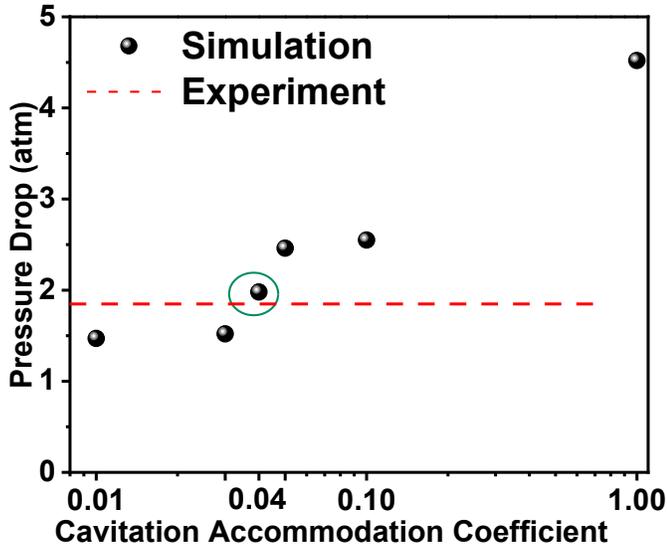
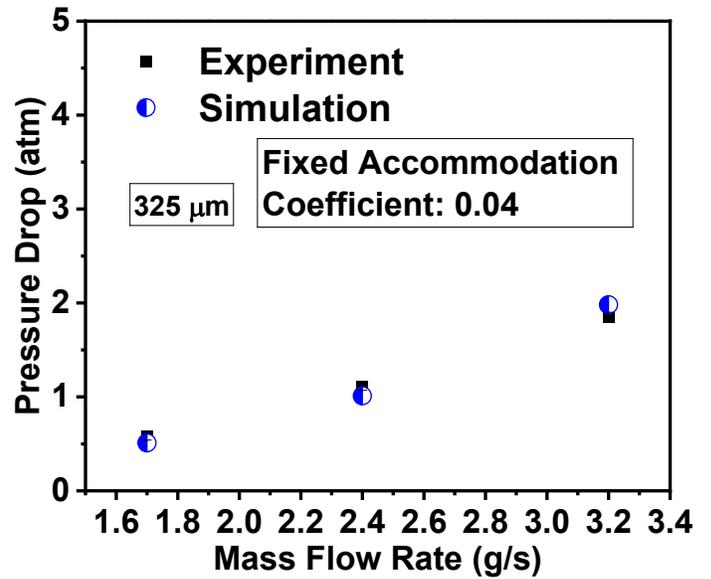
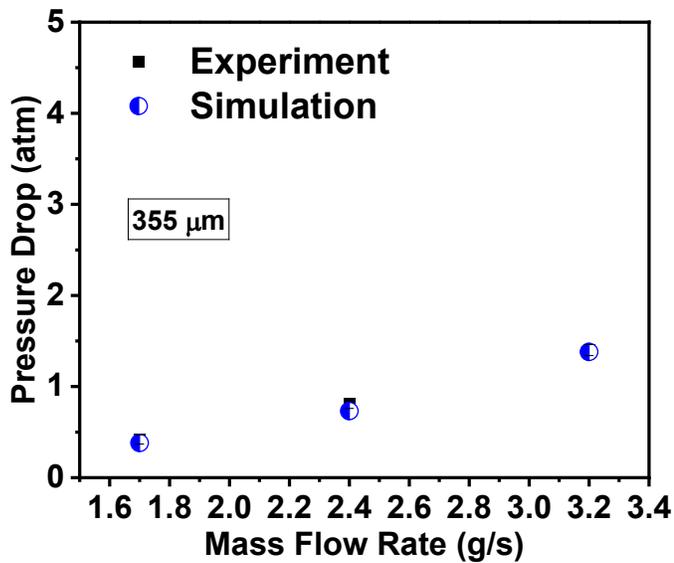
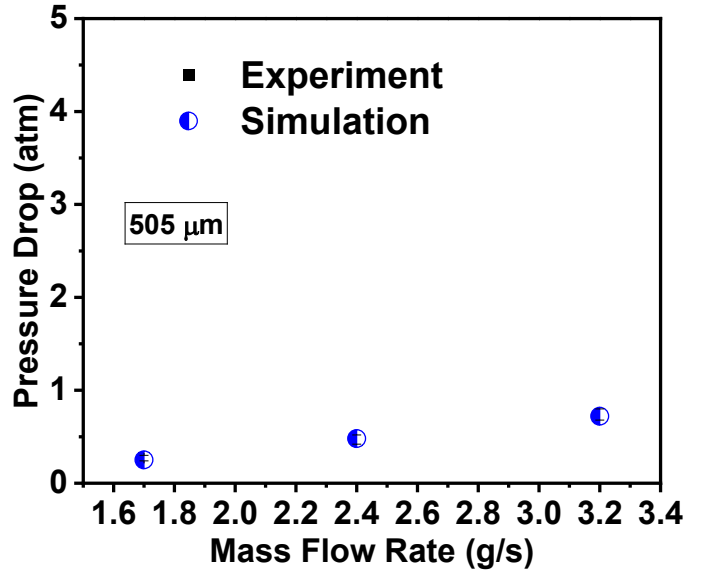

**Figure 6** Model validation using experimental data for the inlet-to-outlet pressure drop in the system. (a) Tuning the cavitation accommodation coefficient for a channel with orifice 325 μm × 300 μm and a fixed mass flow rate of 3.2 g/s. The value of the coefficient that provided best agreement in (a) was used to simulate three additional cases: (b) orifice opening 325 μm × 300 μm, (c) 355 μm × 300 μm, and (d) 505 μm × 300 μm.

The model was then tested with different microchannel geometries across a range of mass flow rates. For the 325 μm × 300 μm orifice, simulations were performed for flow rates between 1.7 g/s and 3.2 g/s, showing excellent agreement with experiments (**Fig. 6b**). The validation was further extended to 355 μm × 300 μm and 505 μm × 300 μm orifice openings, where the mass flow rates also varied from 1.7 g/s to 3.2 g/s (**Figs 6c** and **6d**). Across all cases, the model consistently captured the pressure drop trend observed experimentally, demonstrating its robustness.

### 3.3 Cavitation in the Microchannel

Initially, the temporal evolution of key flow parameters (pressure, velocity and vapor fraction) along the microchannel was examined for a representative case to understand the relevant dynamic characteristics. Subsequently, the influence of the opening angle of the channel on the pressure drop was investigated to assess this angle's impact on flow resistance and potential cavitation zones. The effect of orifice width to prechamber width ratio (β) on channel pressure drop was also investigated. Finally, the effect of varying mass flow rates on the pressure distribution along the channel was studied to evaluate how operational conditions influence the overall pressure drop and flow performance. These analyses provide insights into the interplay between microchannel geometry, flow rates, and cavitation behavior, all critical for optimizing multi-phase cooling system designs.

**3.3.1 Spatio-temporal Evolution of Key Flow Parameters**

The cavitation evolution in the microchannel can be understood by temporally and spatially resolved pressure, velocity, and vapor volume fraction fields. Complete visualizations, including time-resolved pressure maps, velocity–vapor interactions, and mid-plane cross-sectional fields, were performed. The key flow parameters for a microchannel with orifice width 355 μm and mass flow rate of 4.5 g/s (which exceeds the highest mass flow rate considered in the experiments) is presented here for an opening angle of 14º and 30.4 mm channel length. **Figure 7** presents the pressure distribution near the ceiling of the diverging microchannel at four different time instances: (a) 1 ms, (b) 4 ms, (c) 7 ms, and (d) 10 ms from the onset of the flow. A persistent low-pressure zone is visible near the orifice in all cases, highlighted by the blue region. This zone corresponds to the geometric constriction, which accelerates the flow and induces a localized pressure drop. Although the simulation is transient, the overall pressure distribution pattern remains largely unchanged for all timesteps. The position and intensity of the low-pressure zone do not significantly change, suggesting that the pressure field stabilizes rapidly after flow initiation. This indicates that the flow approaches a quasi-steady state during the observed time window, with the cavitation-prone region remaining confined to the orifice vicinity.

**Figure 8** illustrates the transient velocity distribution across the microchannel domain (on a plane at a distance of 20 μm from the ceiling) at the same four instances, i.e., 1 ms, 4 ms, 7 ms, and 10 ms. Early on, the velocity field is smooth and uniform, with the highest velocities (~45 m/s) concentrated near the divergent channel inlet and gradually decreasing downstream, indicating a stable and streamlined flow. By 4 ms, signs of flow disturbances begin to appear downstream as cavitation takes hold, leading to localized fluctuations in velocity. At 7 ms, the velocity field becomes increasingly disrupted, with moderate to low-velocity regions forming in response to growing vapor structures and pressure perturbations. By 10 ms, the flow exhibits significant unsteadiness and turbulence, with distinct low-velocity pockets and vortex-like structures aligning with known cavitation zones. These velocity changes correlate strongly with the pressure drop and vapor volume fraction trends (below), confirming that cavitation significantly influences the turbulent flow dynamics in these diverging microchannels.

**Figure 9** illustrates the cross-sectional velocity map at the orifice of the microchannel, with the mesh grid superimposed for reference. A shallow zero-velocity region is observed along the channel side walls, a consequence of the no-slip boundary condition. Moving inward from the boundaries, a noticeable high-velocity zone develops near the side walls. This phenomenon is attributed to a nozzle-like acceleration effect arising from the curved geometry and narrow spacing at the channel's entry region. The confinement created by the closely spaced side walls focuses the flow, intensifying the local streamwise velocity and producing a jet-like feature adjacent to the walls.

**Figure 10** presents the transient evolution of vapor volume fraction across the microchannel domain (on a plane at a distance of 20 μm from the ceiling) capturing the dynamic cavitation behavior at the same four time instances. Early on, minimal vapor generation is observed near the inlet, with localized vapor volume fractions near the ceiling, indicating the early onset of cavitation due to rapid pressure drops. As the flow progresses, the 4 ms vapor volume fraction map shows that vapor accumulation intensifies, forming distinct vapor pockets (vapor volume fraction approaches unity) spreading from the ceiling towards the floor, signifying the growth of active cavitation zones. These vapor pockets are convected further downstream, following the regions of lowest pressure within the channel. By 7 ms, cavitation is fully developed, with large, continuous (floor-to-ceiling) vapor structures dominating the core flow region, evidenced by extensive red zones representing near-complete phase change from liquid to vapor. This later stage highlights the aggressive nature of cavitation under sustained low-pressure local conditions. At 10 ms, the vapor structures persist but show signs of redistribution, with some vapor collapse near the inlet and a downstream shift of the primary cavitation zones. The detailed zoomed views emphasize the localized nature of bubble clusters and the complex interactions in the turbulent flow.

Overall, Figs 7-10 underscore how geometric divergence and flow conditions critically influence vapor formation —factors that are vital in predicting cavitation risks in microchannel-based cooling systems for electronics.

**Figure 11** maps the vapor volume fraction distribution along vertical cross-sections along the centerline of the diverging microchannel at different time instances, complemented by velocity vector fields that help rationalize the flow behavior. This sectional view offers a more detailed understanding of how vapor structures evolve not only along the length of the channel but also across its height. At 1 ms, the vapor volume fraction remains low and localized near the upper portion of the channel wall, suggesting that initial nucleation of vapor bubbles occurs at the upper boundary layer due to local pressure drops. As time progresses, the zoomed-in regions clearly reveal that vapor regions grow in size and intensity, gradually occupying more of the channel cross section. By 4 ms and 7 ms, the vapor volume fraction significantly increases, with the central core of the channel showing elongated vapor pockets and strong asymmetry in distribution. This uneven distribution suggests that cavitation is not uniform across the height of the channel and is heavily influenced by local velocity gradients and shear layers. The velocity vectors also show the development of complex flow patterns, such as recirculation and vortex motion near the vapor zones, indicating strong coupling between cavitation and flow field dynamics. At 10 ms, well-developed vapor pockets appear and span a large portion of the channel height, with dense vapor regions exhibiting volume fractions approaching 0.9, primarily concentrated around the core and upper midsection of the channel. These maps affirm that cavitation evolves in both the stream-wise and cross-stream directions, with boundary interactions and shear-driven effects playing a significant role in vapor bubble growth and vapor pocket migration within the microchannel.

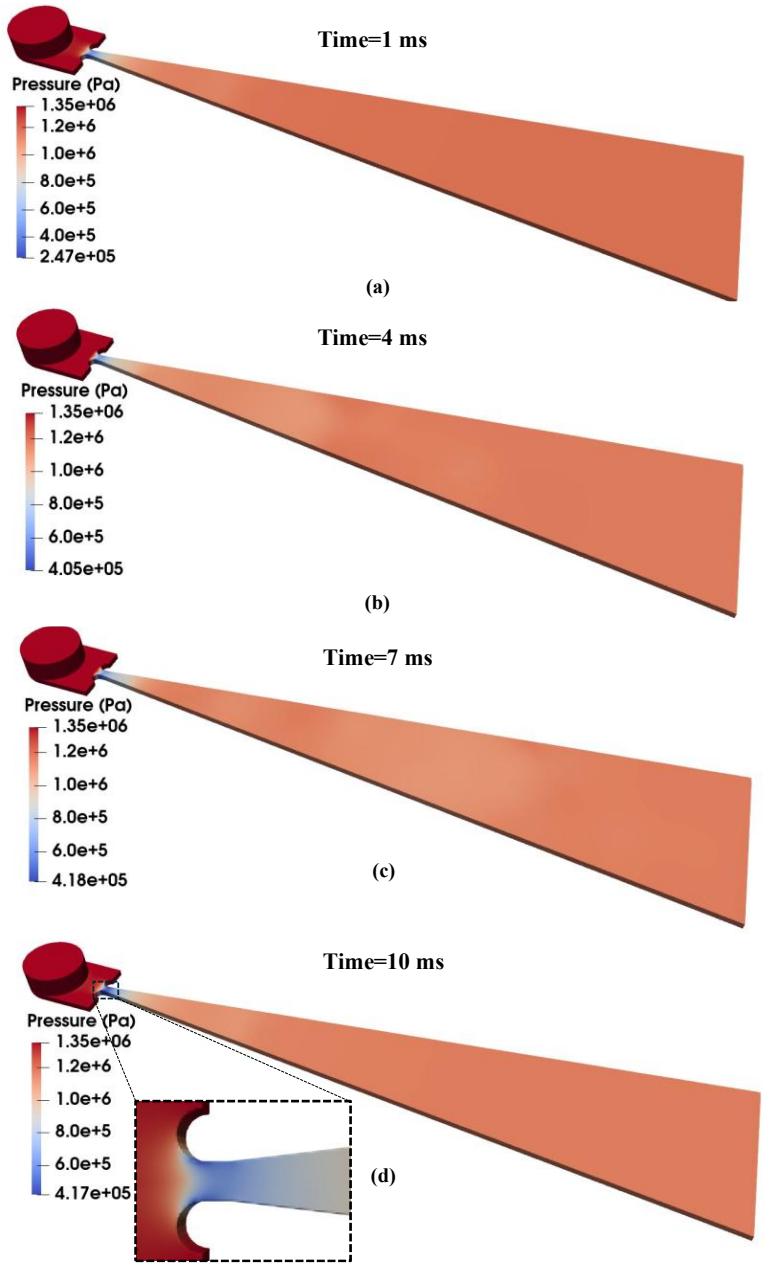

**Figure 7.** Spatial maps of pressure across the microchannel domain (on a horizontal plane at a distance of 20 μm from the ceiling) with orifice width 355 μm, length 30.4 mm and a high flow rate 4.5g/s at: (a) 1ms, (b) 4ms, (c) 7ms, and (d) 10ms from flow onset. The orifice constricts the flow, raising the velocity, and lowering the pressure. Pressure recovery can be seen at the downstream portions of the divergent channel. The inset in (d) presents a magnified detail of the orifice region.

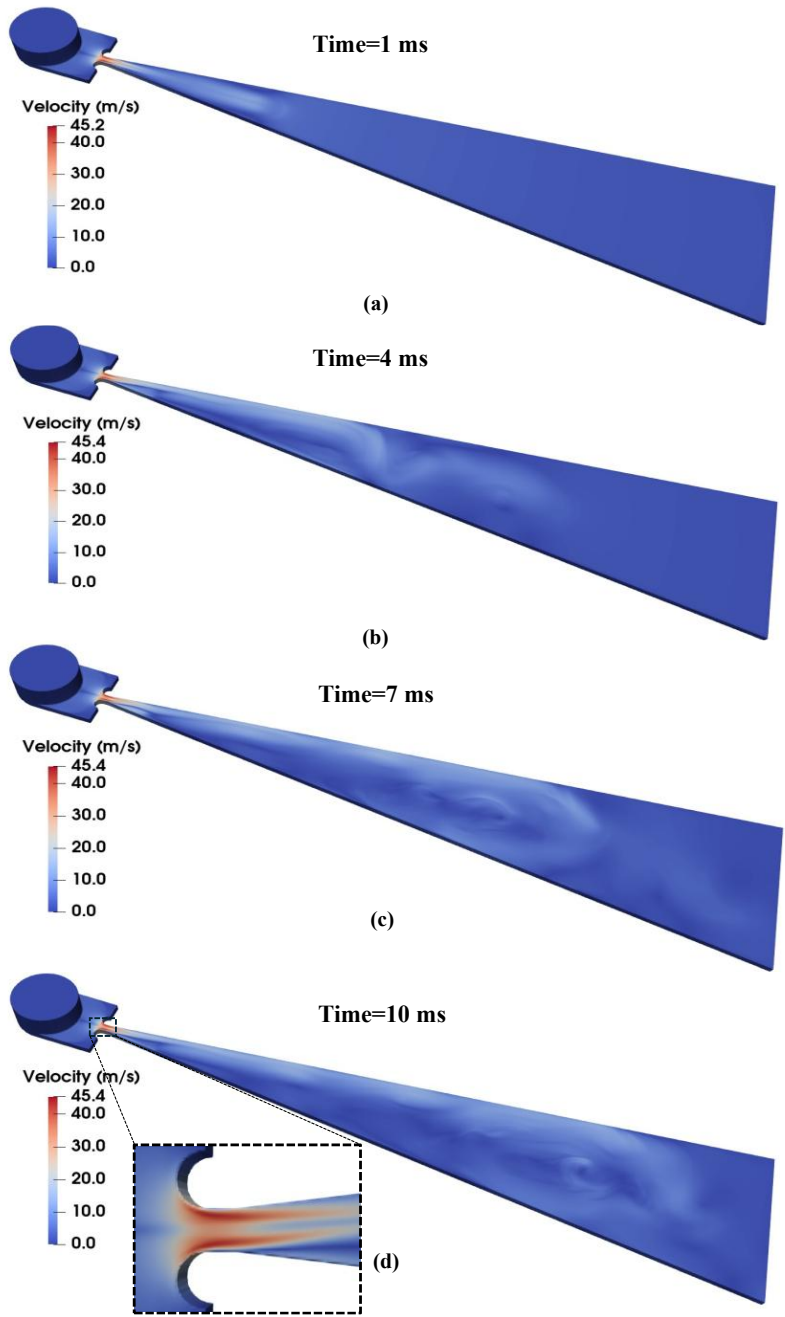

**Figure 8.** Spatial maps of velocity magnitude across the microchannel domain (on a horizontal plane at a distance of 20 μm from the ceiling) with orifice width 355 μm, channel length 30.4 mm and a high flow rate 4.5g/s at: (a) 1ms, (b) 4ms, (c) 7ms, and (d) 10ms from flow onset. The orifice constricts the flow, raising the velocity, and creating vortical structures that are "washed" downstream. The inset in (d) presents a magnified detail of the orifice region and reveals two high-velocity vertical bands near the sidewalls.

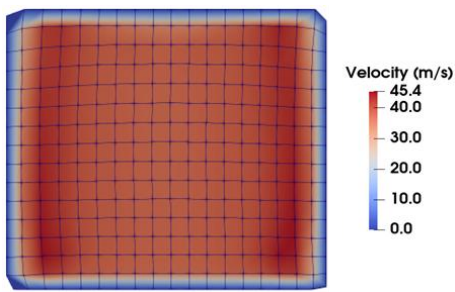

**Figure 9.** Cross-sectional velocity magnitude map at the orifice of the channel with superimposed mesh distribution in this domain. The two higher-velocity vertical bands can be seen near the sidewalls.

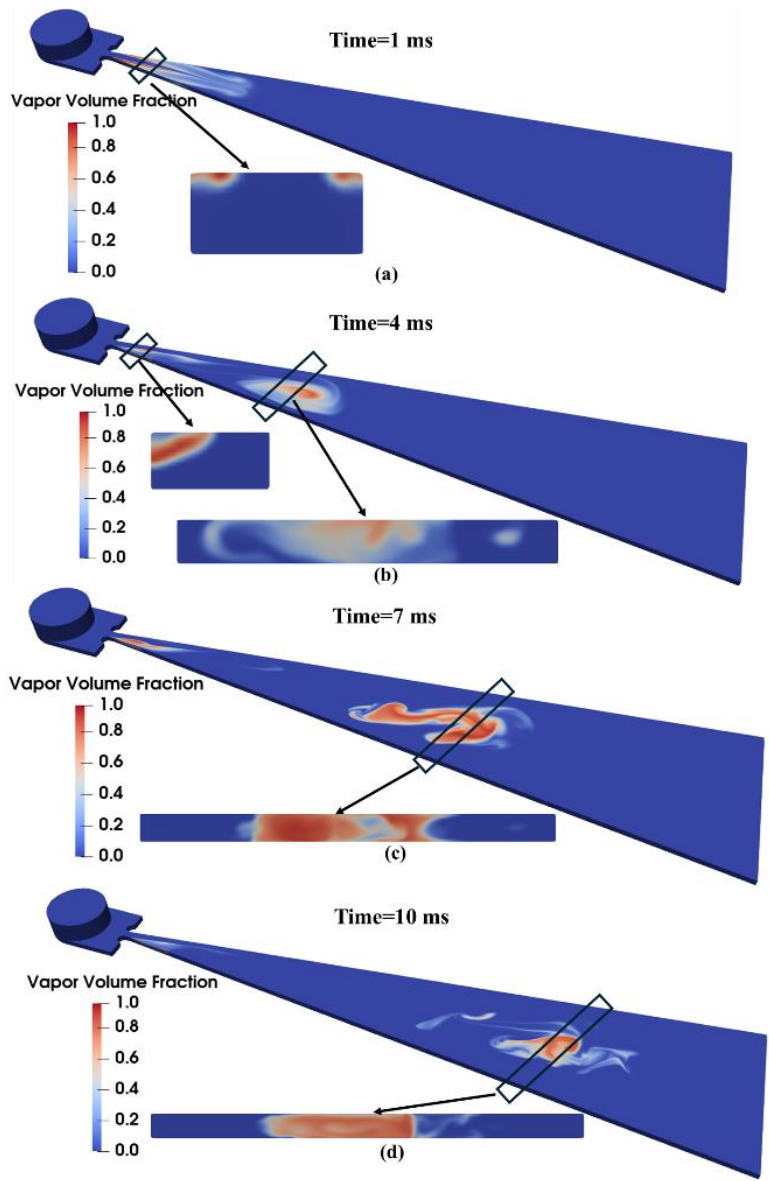

**Figure 10.** Spatial maps of vapor volume fraction across the microchannel domain (on a horizontal plane at a distance of 20 μm from the ceiling) with orifice width 355 μm, channel length 30.4 mm and a flow rate 4.5g/s at: (a) 1ms, (b) 4ms, (c) 7ms, and (d) 10ms from flow onset. The orifice constricts the flow, raising the velocity, and lowering the pressure, which causes cavitation and the formation of vapor pockets that get "washed" downstream. The rectangular color fields show the vapor distribution throughout select cross sections marked with black rectangles along the channel.

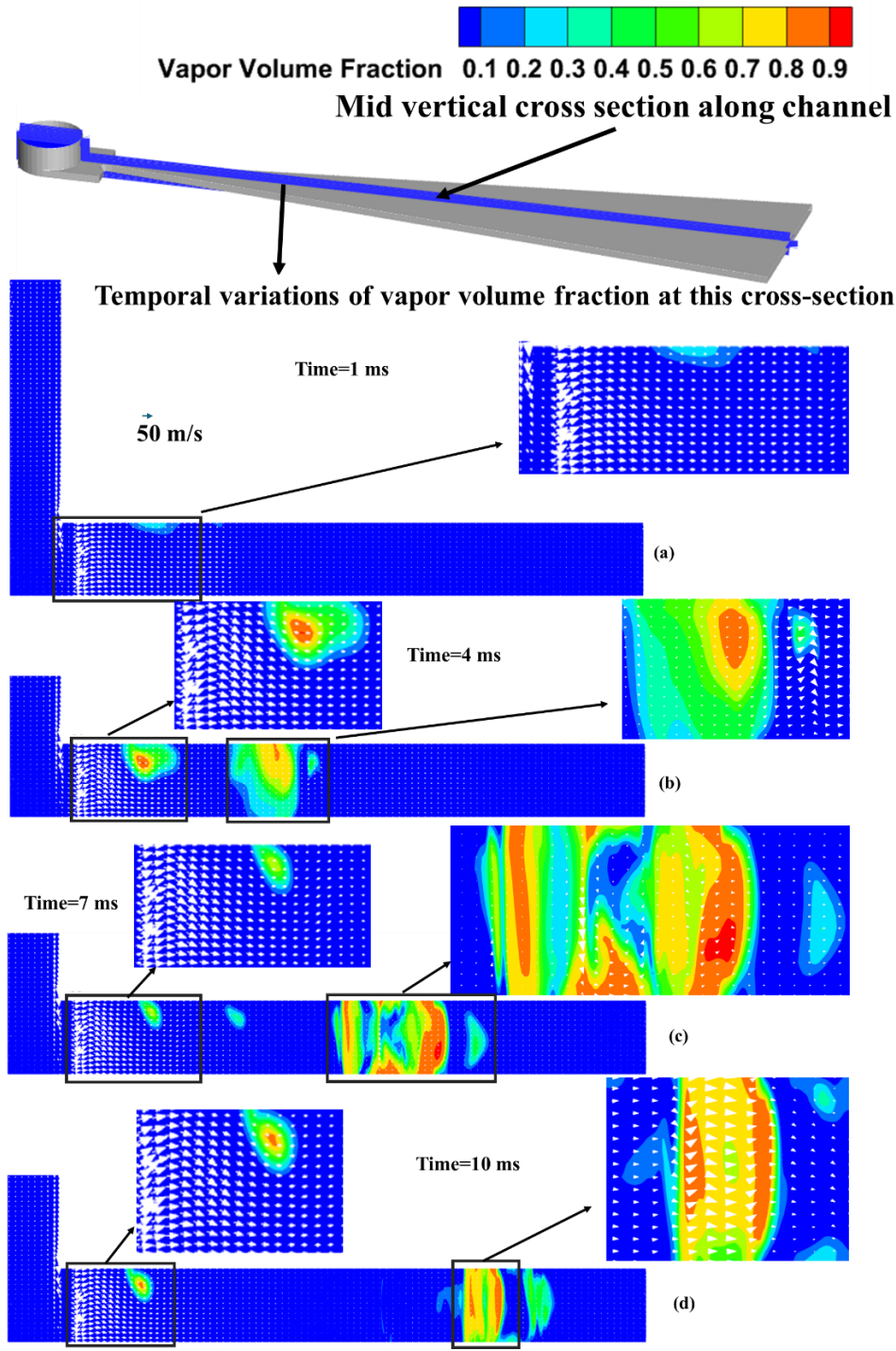

**Figure 11.** (Top) Depiction of vertical cross section along the lengthwise axis of the system. Overlaid side views of velocity and vapor volume fraction fields on the vertical cross section at: (a) 1ms, (b) 4ms, (c) 7ms, and (d) 10ms from flow onset. The spatial scale along the channel height has been magnified to reveal flow structure detail. The orifice constricts the flow, raising the velocity there. The vapor volume fraction fields show the distribution of vapor within the flow domain, which is modulated by local velocity gradients driving the spatial migration of the vapor pockets.

The simulations reveal a pronounced low-pressure region forming near the orifice immediately after flow onset due to geometric constriction, initiating localized cavitation. This pressure minimum remains persistent while the downstream pressure gradually recovers and stabilizes, reaching a quasi-steady distribution beyond 10 ms. The velocity field evolves from a smooth, symmetric profile early on, to increasingly distorted turbulent patterns as vapor cavities form. Peak velocities are observed near the orifice in the early stages, but as cavitation intensifies, vapor pockets disrupt the flow core, inducing vortices, recirculation zones, and localized shear layers. These features become more dominant over time, demonstrating the strong feedback between two-phase flow structures and momentum transport. The vapor volume fraction maps reveal a clear progression from localized nucleation at the ceiling-attached boundary layer to the formation of large, asymmetric vapor pockets spanning the entire channel height. By 10 ms after flow initiation, vapor accumulation becomes substantial in the mid-channel region, and influences both the local flow field and the overall flow stability. Cross-sectional profiles further illustrate vapor stratification, with enhanced vapor growth along the ceiling and around the core, consistent with pressure and velocity field evolution.

### 3.3.2 Pressure Distribution Along the Channel

**Figure 12** presents the pressure distribution along the centerline at mid-height (150 μm from ceiling/floor) of the microchannels with orifice widths of 325 μm, 336 μm, 355 μm, and 397 μm with opening angle 14º and four mass flow rates: 1.7, 2.4, 3.2, and 4.5 g/s. In the previous sections, we examined the evolution of velocity fields, pressure contours, and vapor volume fraction distributions along the channel, offering insight into flow behavior and cavitation onset. Here, we focus specifically on the detailed pressure variation along the channel *centerline* for each case to provide a clearer picture of how geometric and flow parameters influence cavitation dynamics. As seen in **Fig. 12** for all configurations, a sharp pressure drop is evident near the channel inlet due to the flow acceleration through the orifice and the associated onset of cavitation. Downstream of this region, the pressure recovers to a level designated by the orifice width and the flow rate. In **Figs. 12a** and **12b** (325 μm and 336 μm orifice widths, respectively), the recovered pressure remains relatively smooth at lower mass flow rates but shows some fluctuations at and above 3.2 g/s. These oscillations point to the formation and collapse of vapor clouds that interact with swirling turbulent structures. This interaction leads to local pressure instabilities, corroborating the earlier observations of high vapor volume fraction and unsteady velocity fields, and highlighting the strong coupling between cavity dynamics and pressure variation.

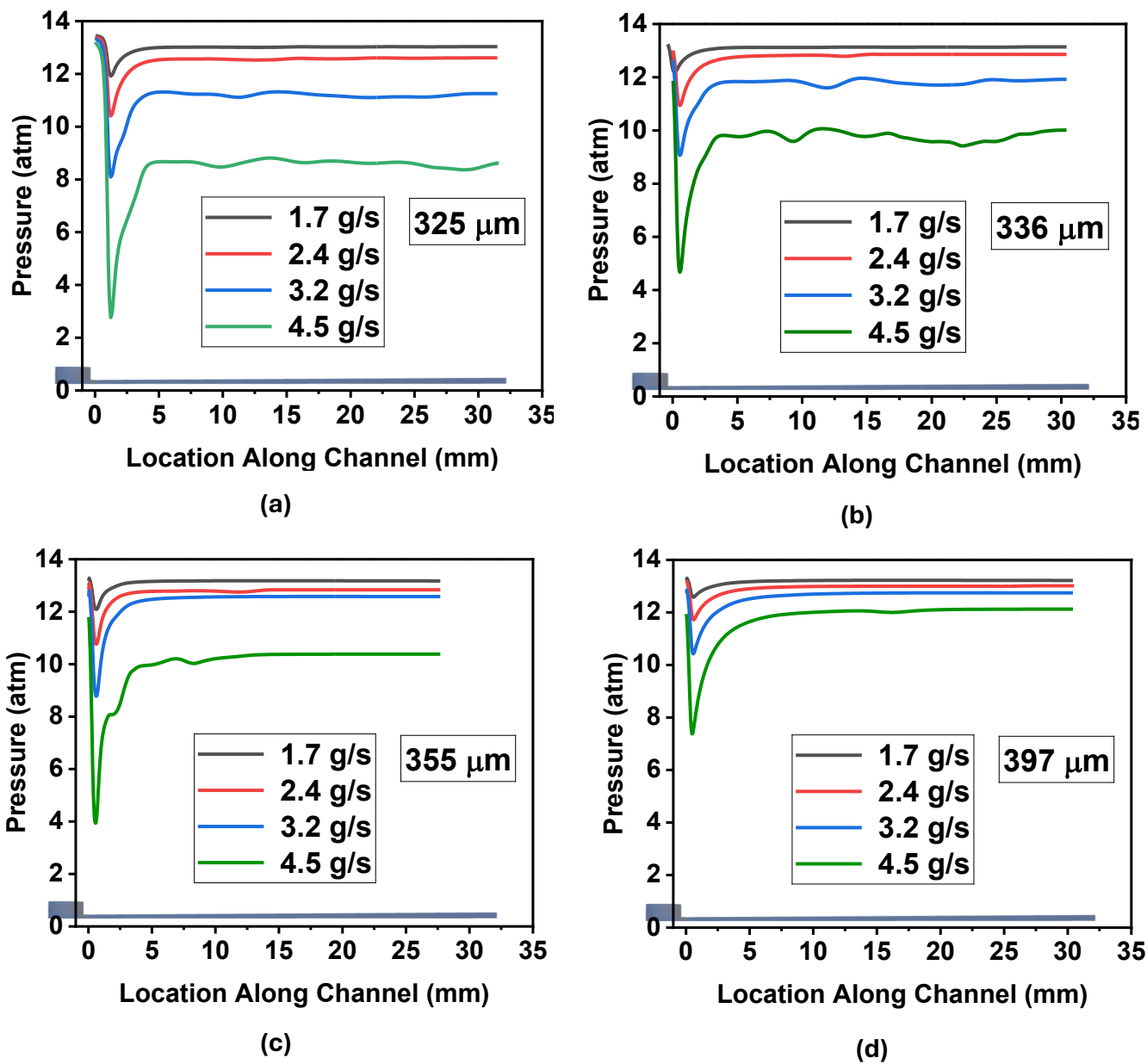

**Figure 12:** Channel centerline pressure variation for different orifice widths: (a) 325 μm, (b) 336 μm, (c) 355 μm, and (d) 397 μm. The opening angle is 14º for these four geometries. A side schematic of the pre-chamber and the channel is shown in grey at the lower left of each graph to correlate the curves with physical location.

As the channel orifice gets wider, the pressure gradient softens and delays or reduces cavitation, consistent with the reduced vapor activity observed previously. In the 355 μm orifice (**Fig. 12c**), the recovered pressure becomes notably stable at 3.2 g/s, with visible oscillations along the channel persisting at 4.5 g/s. Finally, **Fig. 12d** (397 μm) presents the most stable pressure distributions. This widest orifice allows smoother flow expansion, reducing abrupt pressure drops and effectively suppressing cavitation. Even at the highest flow rate, the centerline pressure profile remains steady, highlighting the geometry's effectiveness in maintaining stable flow conditions.

Overall, **Fig. 12** supports the conclusion that narrower orifices combined with higher flow rates amplify pressure instabilities, while wider throat constrictions favor more uniform and predictable pressure behavior—an essential consideration in designing microchannel systems for robust thermal management. The pressure distribution plots (**Fig. 12**) highlight how the throat geometry strongly influences local pressure fields. Smaller openings intensify the throat contraction, leading to sharp depressurization and extended recovery zones. In contrast, larger openings sustain relatively higher downstream pressures and exhibit faster pressure stabilization.

### 3.3.3 Cavitation Number Considerations

The cavitation number serves as a quantitative measure of the cavitation propensity in the channel [17]. It is defined as

$$\sigma = 2\frac{(p_{exit}-p_{sat})d^2H^2}{\rho_l Q^2} \qquad (24)$$

and is plotted in **Fig. 13** as a function of the flow rate. As described in **Fig. 12**, with increasing flow rate, the minimum pressure at the orifice throat drops sharply, resulting in incomplete recovery at the channel outlet and a corresponding reduction in σ across all four geometries. The correlation is evident: the deeper and more sustained the local pressure depression (most pronounced in narrower orifices), the lower the resulting cavitation number. Hence, σ provides a compact -yet powerful- global descriptor of the cavitation propensity that the pressure-field analysis already implies.

Conventionally, the cavitation number is reported as a positive [17], non-dimensional metric derived from the pressure difference between a reference point (typically the outlet) and the vapor pressure, normalized by the dynamic head. Prior studies have considered flow conditions where the fluid pressures exceed the saturation pressure, with the cavitation risk rising as the pressure declines and reaches closer to the saturation value. Under this definition, a decrease in σ denotes an elevated likelihood of cavitation inception. However, in microscale configurations, such as the present orifice-constricted channels where the fluid is supplied at saturated conditions and at high flow rate, the local depressurization can be sufficiently intense that the static pressure falls below the vapor pressure not only at the throat but also at the downstream outlet. Under these conditions, the computed σ can assume negative values—a physically meaningful outcome that signals sustained vapor presence and incomplete pressure recovery. This observation carries important implications. A positive low σ corresponds to incipient or localized cavitation where vapor nuclei appear intermittently; σ → 0 marks the onset of a stable vapor core accompanied by periodic shedding; a negative σ identifies a deeply cavitating regime in which vapor occupies a significant portion of the flow passage and persists even in the diverging section. In this regime, the flow never fully recovers above the saturation limit, indicating continuous vapor connectivity between the orifice and outlet.

Accordingly, both the magnitude and the sign of σ are essential in assessing cavitation severity. The increasingly negative σ values observed here (particularly for the narrowest orifices at the higher flow rates) align with the pronounced pressure deficits and extended vapor zones shown earlier, reinforcing the physical consistency of this metric. While the traditional, strictly positive σ framework suffices for macroscale or moderate-intensity flows, microscale, high-Reynolds-number constricted flows demand explicit consideration of σ's sign. The negative σ thus emerges as an indicator of intensified cavitation and persistent vapor entrapment, refining the classical interpretation of the cavitation number for microscale hydrodynamics.

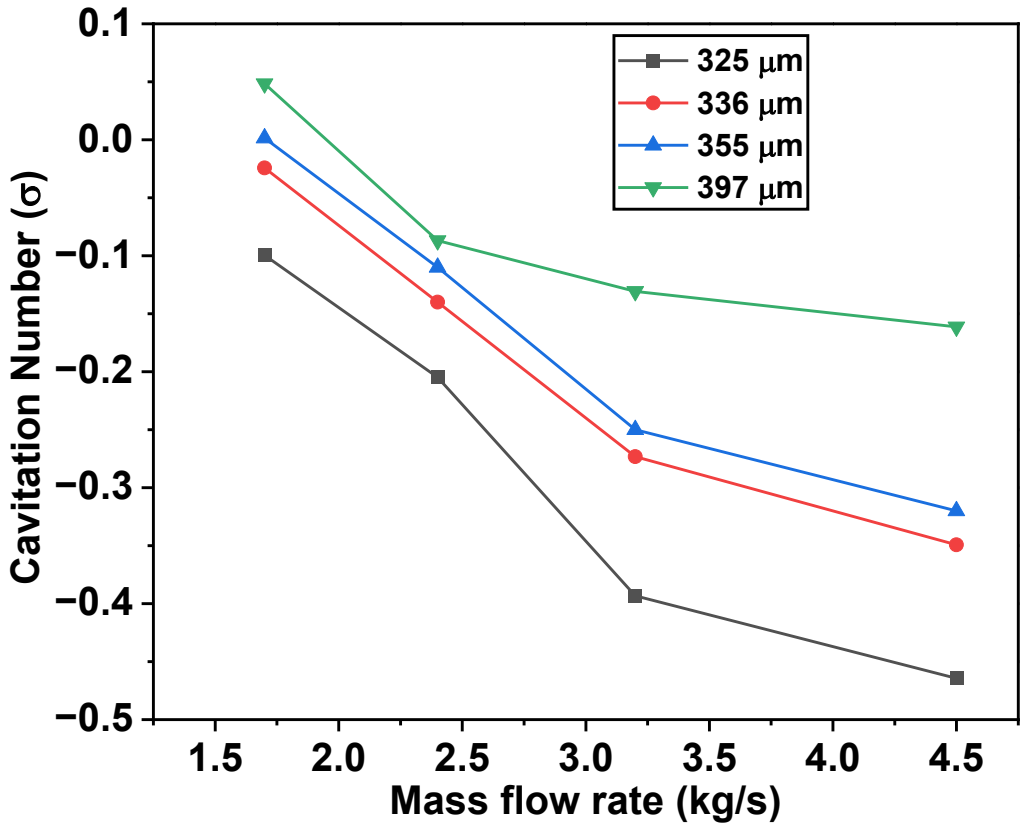

**Figure 13:** Variation of outlet cavitation number with mass flow rate for microchannels of four different orifice openings. For all channels, a consistent decrease in σ is observed with increasing mass flow rate, with narrower orifices exhibiting lower cavitation numbers at the same operating conditions. The strongly negative values of σ indicate a high cavitation risk that is consistent with the experimental observations.

### 3.3.4 Effect of Channel Opening Angle on Pressure Drop

**Figure 14** presents the variation of pressure drop *ΔP* (cross-sectional area averaged values) across a diverging microchannel of length 27.3 mm and orifice dimensions 336 μm × 300 μm vs. opening angle for three different mass flow rates. In all cases, *ΔP* first declines with opening angle, reaches a minimum at around 12°, and then rises for larger angles. At the lowest mass flow rate of 1.7 g/s, the pressure loss is relatively small across all angles, suggesting that the flow remains relatively stable and weakly affected by changes in geometry. As the flow rate increases, the impact of geometry becomes more pronounced. This behavior can be attributed to two competing effects. Initially, increasing the opening angle leads to smoother flow expansion and more efficient pressure recovery, while also suppressing vapor formation via cavitation. Suppression of phase change maintains high mixture density [27] and consequently, low flow velocity (higher pressure), leading to a net reduction in pressure drop. However, beyond a certain angle (~12°), the flow begins to destabilize due to increased expansion and flow separation, resulting in recirculation zones and growing eddies.

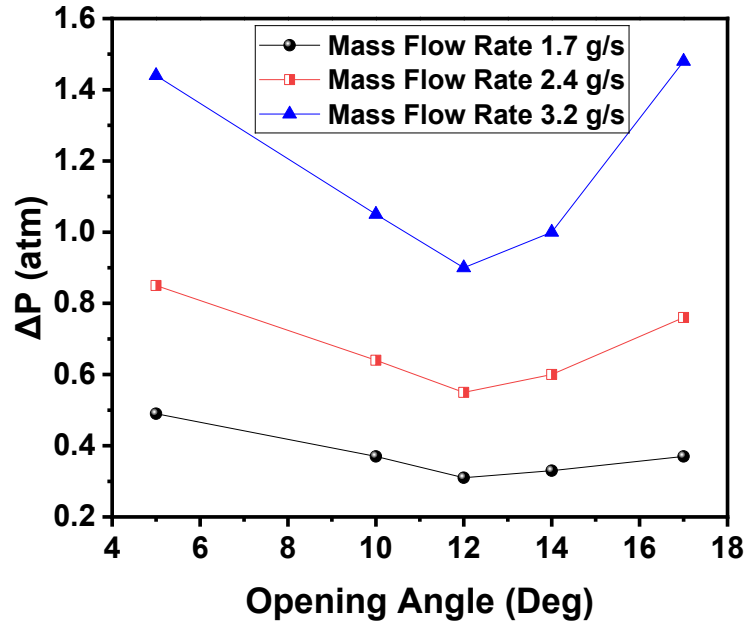

**Figure 14.** Effect of channel divergence angle on pressure drop (cross sectional averaged values) along a channel of length 27.3 mm and orifice dimensions 336 μm × 300 μm.

**Figure 15** illustrates the velocity contours and streamlines across the mid-plane between floor and ceiling for a diverging microchannel of length 27.3 mm and orifice of 336 μm × 300 μm (as in Fig. 13) but with an opening angle of $\theta = 17°$. For this particular case, the mass flow rate is 3.2 g/s. At this higher opening angle, the flow undergoes significant expansion immediately downstream of the orifice, leading to visible recirculation zones along the walls (marked with white-dashed lines) and vortex shedding near the channel core. These separation-induced flow structures disrupt the otherwise streamlined profile, contributing to elevated turbulence intensity and enhanced momentum dissipation, thereby increasing the overall pressure drop across the channel.

The above results illustrate that there exists an optimal opening angle (~12°) where the pressure drop is minimized, especially at higher flow rates that promote cavitation. Beyond this optimal angle, geometric divergence leads to adverse flow effects, including eddies and recirculation zones, which counteract the earlier benefits and elevate pressure losses. The trend is particularly strong at and above 3.2 g/s, indicating that the higher the flow rate, the more sensitive the system becomes to geometric changes. This provides guidance for designing efficient microchannel geometries where minimizing pressure drop is essential for performance and energy efficiency. Such considerations are also critical when one seeks flow balancing across adjacent microchannels that are supplied from the same inlet reservoir; such balancing requires that no channel receives more/less flow than others, an event that can create imbalances in the channels' ability to remove heat via phase change.

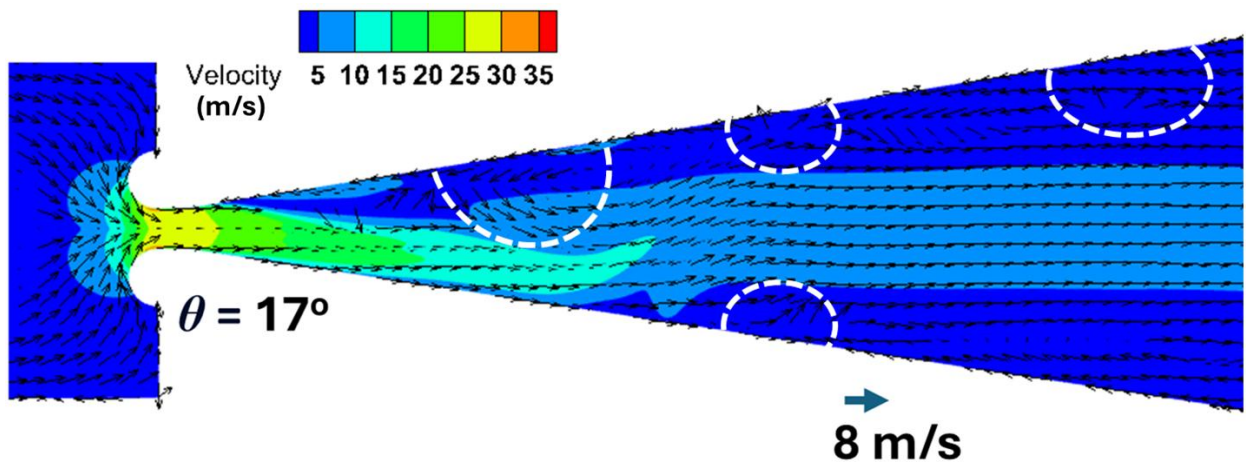

**Figure 15.** Velocity magnitude contours and vectors in a diverging microchannel with an opening angle of 17°. The microchannel has an orifice 336 μm × 300 μm (at left) and length 27.3 mm. The mass flow rate is 3.2 g/s. The values correspond to the mid-plane between floor and ceiling at a distance of 20 μm from the boundary. The flow is asymmetric with respect to the geometric axis, a result of the turbulent character of the flow. The recirculation zones formed near the wall due to high flow rate has been marked with "*dashed white lines*".

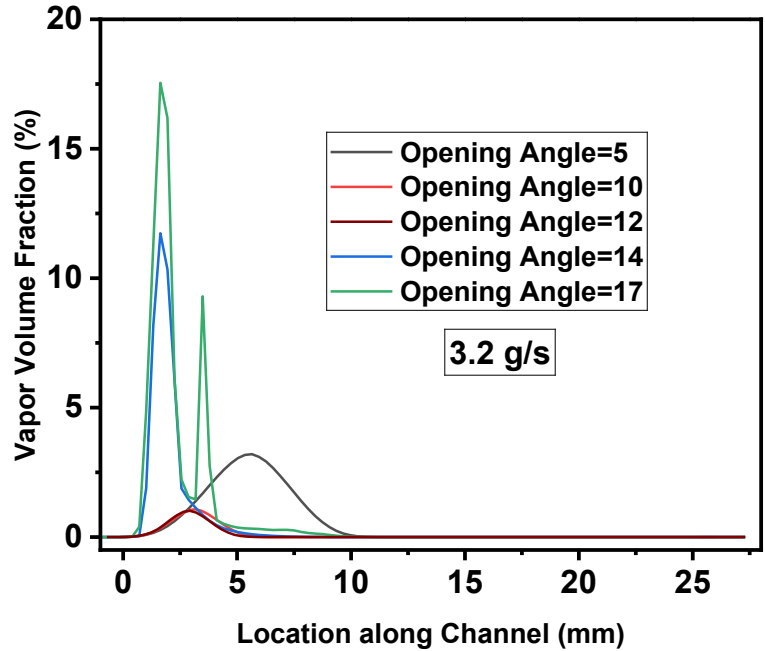

**Figure 16.** Effect of channel divergence angle on cross-section-averaged vapor volume fraction along a channel associated with high cavitation risk (orifice dimensions 336 μm × 300 μm, length 27.3 mm).

**Figure 16** plots the vapor volume fraction (surface-averaged over cross sections normal to the lengthwise axis) along the channel length (27.3 mm) for different opening angles and a constant mass flow rate of 3.2 g/s for an orifice dimension of 336 μm × 300 μm. The data reveals a non-monotonic trend in vapor generation as a function of opening angle. As the opening angle rises from 5° to 12°, the overall vapor generation intensity decreases. The peak vapor volume fraction declines, and the spatial extent of vapor zones becomes narrower along the flow, moving closer to the orifice. This suggests suppression of cavitation, likely due to smoother pressure recovery and reduced local pressure drop along the diverging section. However, beyond 12° opening angle, the trend reverses. For channel opening angles of 14° and 17°, the vapor generation intensifies again, as evidenced by sharp peaks in vapor volume fraction near the inlet region. The reasons for this reversal in behavior and the underlying flow dynamics associated with each stage of this trend will be explained below.

**Figure 17** presents top views of vapor volume fraction and velocity fields at four channel divergence angles for a channel with 27.3 length mm and orifice dimensions 336 μm × 300 μm. A distinct region of high-velocity fluid concentrated near the side walls beyond the flow entry is pronounced at all opening angles. This is attributed to a nozzle-like acceleration effect caused by the curvature and narrow spacing of the entry walls. The phenomenon is analogous to the nozzle effect described by Li et al. [42], where gap constraints between side-by-side bodies intensify the flow in the interstitial region. In the present case, the higher local velocity near the side-walls is due to reduced pressure buildup and enhanced streamwise momentum, which together generate jet-like features tightly hugging the lateral boundaries of the orifice. This behavior highlights the sensitivity of inlet flow dynamics to geometric divergence, with direct implications for heat transfer and pressure loss characteristics in curved duct systems.

31Approved for Public Release: NG25-2102. ©2026 Northrop Grumman Systems Corporation

The emergence of high-velocity domains near the side-walls at the channel inlet lays the foundation for the downstream evolution of the flow field. As the opening angle rises, the same geometry that intensifies the fluid acceleration near the orifice also introduces adverse pressure gradients that gradually destabilize the flow. Consequently, while the inlet region exhibits strong directed momentum, the downstream section transitions from attached flow to separation and vortex formation. This interplay between localized acceleration and global instability underscores the dual role of inlet angle divergence: it first promotes jet formation and then triggers flow separation, as elaborated below.

**Figure 17a** illustrates that for low channel opening angles, vapor generation due to cavitation is primarily observed downstream of the orifice, with the vapor regions formed near the channel ceiling and side-walls. This spatial distribution is influenced by the curvature-induced near-wall acceleration just past the orifice, which generates a localized low-pressure region near the side boundaries, promoting vapor formation along those regions. The corresponding velocity field in Fig. 17(a) shows a highly focused, attached (to the side-walls) flow, with no visible recirculation or eddy structures, indicating a streamlined behavior. The upward shift of vapor toward the ceiling is explained in fig.1; the floor-to-ceiling asymmetry arises from velocity gradients across the channel height. As the channel divergence angle increases, the vapor generation region shifts closer to the orifice, as observed in **Figs. 17(b-d)**. This is due to the decreased flow velocity at higher channel opening angle, which reduces the downstream transport of vapor and confines the cavitation zone nearer to the inlet. For an opening angle of 10° (**Fig. 17b**), the vapor generation intensity is much lower than for 5°, due to the broader channel entry resulting in lower velocities and milder pressure gradients. In other words, the system behavior improves from a cavitation perspective, as evidenced by the lower vapor volume fractions in **Fig.**

**17b.** At higher opening angles, **Figs. 17c** and **17d**, the flow behavior changes markedly. The vector plots (at bottom) show clear evidence of flow separation and the formation of eddies. At 14°, flow separation along the diverging walls generates recirculation zones. These recirculation structures modify the pressure distribution and assist in detaching the vapor core from the side walls. By 17°, these eddies become more pronounced and widespread, occupying a significant portion of the diverging channel. These recirculation zones contribute to increased local pressure at these high angles, as indicated in **Fig. 15**. Furthermore, with the increase in flow separation and emergence of recirculation zones, local pockets of low-pressure regions are created. These localized low-pressure pockets amplify vapor formation, which ultimately leads to increased pressure losses, as depicted in **Fig. 14**.

In summary, the pressure loss variation with channel divergence angle and flow rate is the outcome of two competing mechanisms: (a) cavitation, and (b) flow separation. At low to moderate divergence angles (up to approximately 12°), cavitation is the dominant phenomenon due to the spatially constricted flow near the orifice. Cavitation weakens as the channel angle rises, and thus the associated pressure losses also decline. For channel divergence angles beyond 12°, the influence of cavitation weakens and flow separation and recirculation effects become more pronounced. These structures contribute to increased energy dissipation, flow instability, and ultimately, greater pressure losses.

These findings emphasize the importance of careful optimization of the microchannel opening angle in coordination with the orifice dimensions. For the present geometry, an intermediate angle around 12° offers an optimal configuration—promoting controlled cavitation for pressure reduction, while avoiding the onset of flow separation and energy loss. The integrated analysis of vapor volume fraction, velocity magnitude and vector fields across different geometries promotes the understanding of the underlying flow physics, supporting the conclusion that microchannel performance is strongly governed by the interplay between geometry-driven flow stability and cavitation dynamics.

33Approved for Public Release: NG25-2102. ©2026 Northrop Grumman Systems Corporation

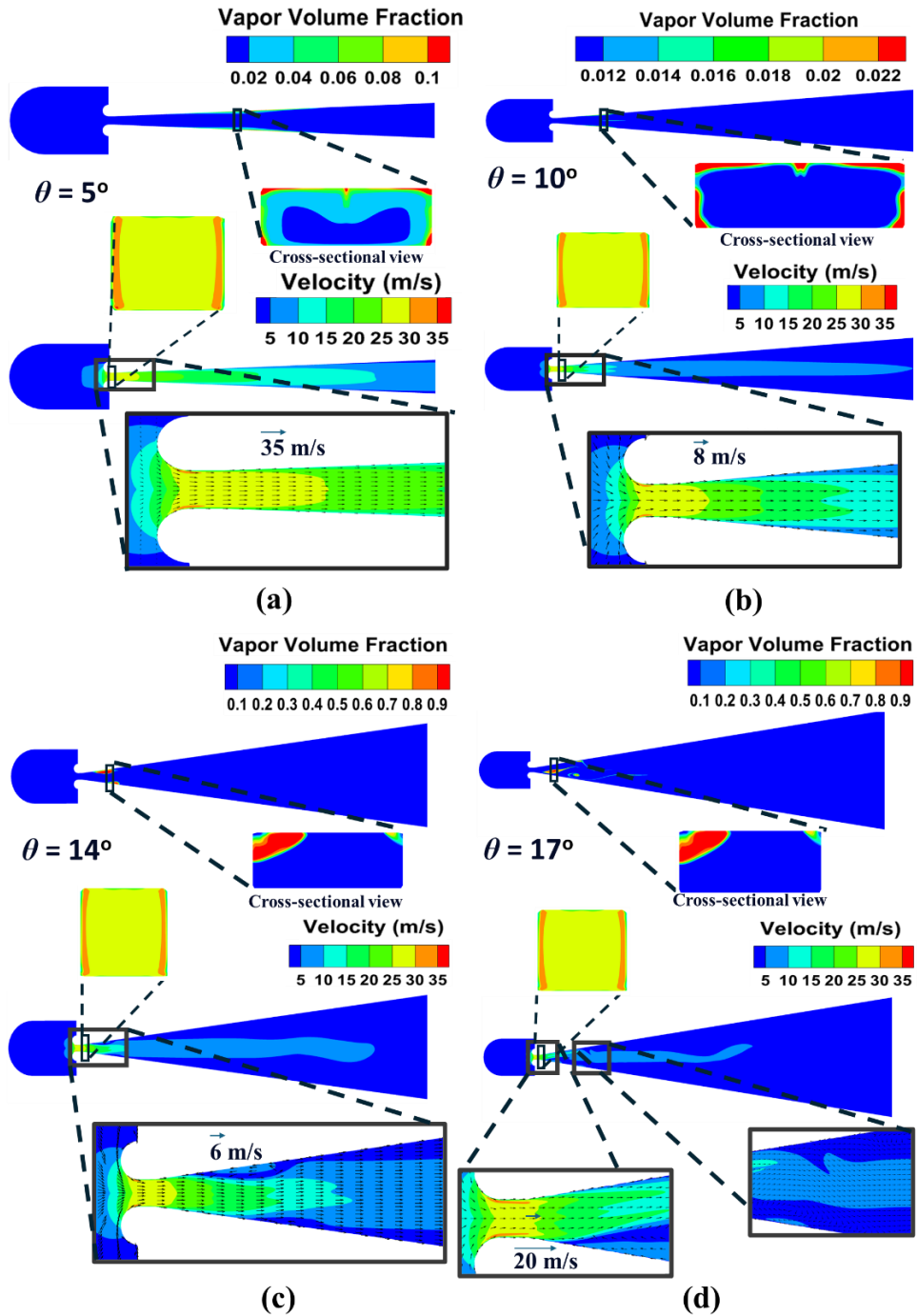

**Figure 17.** Distribution of vapor across the channel mid plane and velocity vectors juxtaposed with velocity contour maps for flow across different parts of the channel with opening angles of: (a) 5º, (b) 10º, (c) 14º, and (d) 17º. The microchannel has a length of 27.3 mm and orifice dimensions 336 μm × 300 μm. The mass flow rate is 3.2 g/s. Selected cross sections are shown with velocity maps close to the orifice and vapor volume faction contours, revealing the spatial distribution of the vapor phase in these areas.

### 3.3.5 Effect of Orifice-to-Prechamber Width Ratio on Pressure Losses

**Figure 18** illustrates the effect of β on the pressure losses (cross sectional averaged values) along different microchannels of length 27.3 mm and opening angle of 14º. Both simulation and experimental ΔP data are plotted and compared. In all cases, the pre-chamber width *D* was fixed, but the orifice width *d* varied, thus changing the value of β. **Figure 18a** presents results for the lower mass flow rates (1.7 g/s, 2.4 g/s), while **Fig. 18b** plots data for the higher flow rates (3.2 g/s, 4.5 g/s). Across all flow conditions, the pressure drop declines as the orifice widens (β rises). Wider orifices relative to the pre-chamber (high β) allow a smoother flow transition from the pre-chamber to the orifice, reducing the intensity of local acceleration, turbulence, and vapor bubble formation. As a result, for these cases the overall pressure loss through the microchannel is reduced. Narrower orifices (low β) cause a more abrupt contraction of the flow, leading to higher local velocities, steep pressure gradients, and increased likelihood of cavitation and flow separation. These effects contribute to higher flow resistance and greater pressure loss. The trends are well captured by both the simulation and experimental results. At all flow rates, the agreement between simulation and experiment is excellent, with minimal deviation and consistent trends. In summary, **Fig. 18** confirms that increasing the β ratio is an effective strategy for reducing overall pressure losses in diverging microchannels. This is important for optimizing microchannel design in applications where managing pressure drop and cavitation is essential for ensuring system performance and longevity.

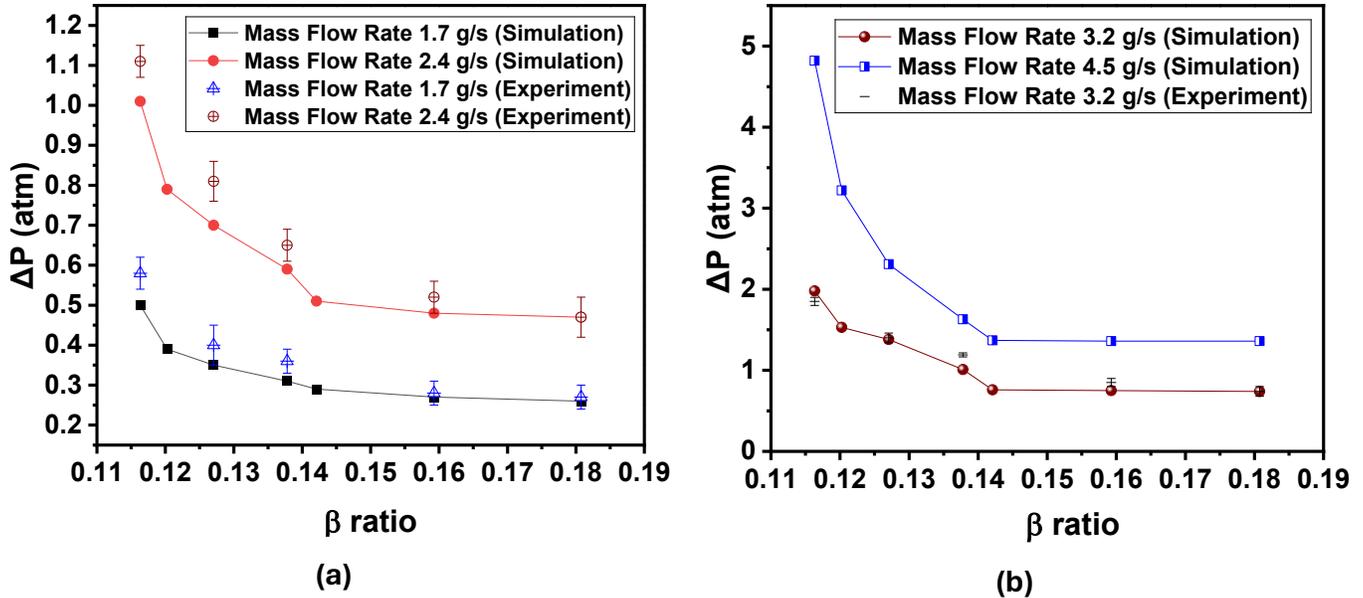

**Figure 18.** Effect of β ratio on pressure losses along channels of length 27.3 mm and opening angle of 14º but having the same pre-chamber but different orifice widths. Both numerical and experimental data are plotted for four flow rates. No experimental data were obtained for 4.5 g/s.

### 3.3.6 Effect of Mass Flow Rate on Pressure Losses

**Figure 19** presents the effect of mass flow rate on *ΔP* along microchannels of different orifice widths (*d*); both experimental data and simulation results are plotted. The orifice widths ranged from 325 μm to 505 μm, the channel length was fixed at 30.4 mm for all cases, and the opening angle was 14°, thus resulting in varying channel exit widths. The figure demonstrates how both flow rate and orifice geometry jointly influence hydraulic resistance and pressure losses in the system. The overall trend is consistent: pressure losses rise with mass flow rate for all orifice sizes. However, the magnitude of the pressure drop and its rate of increase are highly dependent on the orifice width. For narrower orifices, the pressure drop rises steeply with flow rate. In contrast, for wider orifices, the pressure drop rises more gradually and remains significantly lower across all flow rates. This is attributed to narrower orifices causing higher velocity acceleration and greater local pressure gradients, particularly at higher flow rates. These effects promote cavitation onset, turbulence, and energy loss, all of which contribute to increased pressure losses. On the contrary, wider orifices allow for smoother expansion and lower velocity peaks, which help suppress cavitation and reduce hydraulic resistance.

The qualitative and quantitative agreement between experiment and simulation in **Fig. 19** is strong, with both datasets revealing two key outcomes. First, mass flow rate has a nonlinear and geometry-dependent effect on pressure losses, with narrower orifices amplifying pressure losses due to enhanced acceleration and cavitation effects. Second, the simulation results are consistent with experimental observations, offering valuable predictive capability for evaluating pressure behavior across a range of flow and geometric conditions. These insights are critical for designing high-performance microchannel systems where minimizing pressure drop is essential for efficient thermal regulation or balancing flows across multiple channels.

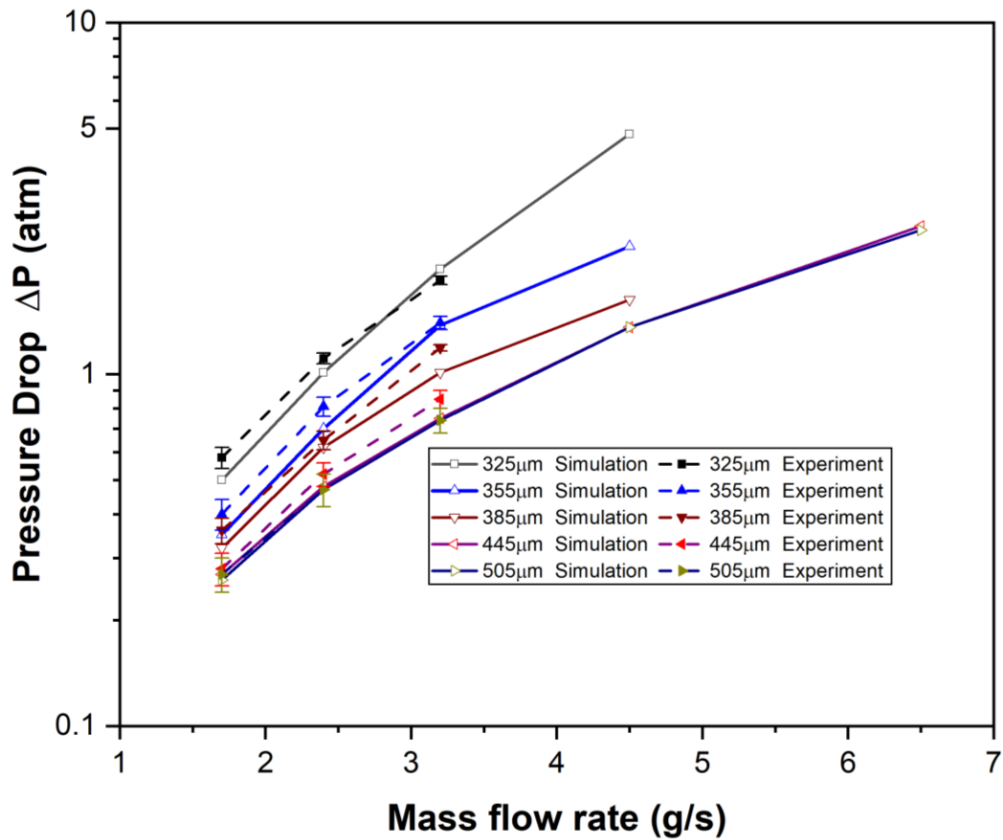

**Figure 19.** Effect of mass flow rate on pressure drop along channels of varying orifice widths. Both experimental and modeling results are plotted. Channel length is 30.4 mm and opening angle is 14 °.

## 4  Conclusions

Based on a comprehensive analysis, a computational framework was developed to predict cavitation-induced flows for R134a refrigerant in divergent microchannels using a Volume of Fluid (VOF) approach coupled with the Schnerr-Sauer cavitation model. The channel orifice width was in the range 325-505 μm, the channel height was fixed at 300 μm, and the channel lengths varied in the range 27.3-30.4 mm, resulting in channel opening angles from 5º to 17º. The cavitation model was calibrated initially using experimental data for a divergent channel featuring a 325 μm-wide orifice and supplied at a mass flow rate of 3.2 g/s. The model's predictive accuracy was further confirmed experimentally across a range of orifice sizes and refrigerant flow rates, demonstrating robustness across varying geometries.

Key flow characteristics, such as pressure, velocity, and vapor volume fraction were investigated and mapped. The results revealed that pressure loss is governed by two competing mechanisms, namely vapor generation due to cavitation (which rises at narrower orifice widths for fixed flow rate) and flow separation (which increases at higher opening angles). Optimal performance (lowest pressure loss) was consistently observed around 12° opening angle, where the balance between the two competing mechanisms is most favorable. Additionally, increasing the orifice-to-prechamber width ratio reduced pressure drop by enabling smoother flow transitions and mitigating the onset of cavitation. The results showed that pressure losses increase with mass flow rate for all orifice sizes, with narrower orifices exhibiting steeper pressure losses due to higher velocity gradients and increased cavitation intensity. In contrast, wider orifices produced more even pressure fields and lower overall resistance.

Analysis of the outlet cavitation number σ revealed that as the mass flow rate increases, throat pressure drops more sharply, leading to reduced outlet pressure recovery and progressively lower σ values. Notably, for high-Reynolds-number microscale flows through severe constrictions, σ may even become negative, indicating sustained vapor presence and incomplete pressure recovery downstream. Consequently, both the magnitude and the sign of σ serve as crucial quantitative indicators of cavitation severity—where increasingly negative values correspond to stronger vapor persistence and higher cavitation risk at and beyond the orifice.

The validated computational model developed in this work serves as an effective design and diagnostic tool for high-speed microscale flow applications, including microfluidic chemical synthesis, biomedical diagnostics, lab-on-chip platforms, and compact energy systems where cavitation and multiphase effects play a significant role. The results provide important insights into how microchannel geometry and flow conditions influence cavitation behavior and associated pressure losses, thereby guiding the development and optimization of advanced microfluidic systems for diverse applications in science and technology.

**Acknowledgements**

The authors acknowledge financial support from the Defense Advanced Research Projects Agency (DARPA) under contract HR001124C0303 to Northrop Grumman Corporation. Any opinions, findings and conclusions or recommendations expressed in this material are those of the authors and do not necessarily reflect the views of DARPA. The simulations were carried out using the computational resources provided by the National Center for Supercomputing Applications (NCSA) through the DELTA cluster high-performance computing facility. The authors gratefully acknowledge the support of NCSA for enabling the numerical simulations conducted in this work. The authors extend their gratitude to Karl Frietag and Carlos Herrera of Northrop Grumman Corporation for their assistance in flow loop construction and instrumentation/control integration. MATLAB® is a registered trademark of MathWorks, Inc. OpenFOAM® is the free, open source CFD software developed primarily by OpenCFD Ltd since 2004. DWYEROMEGA$^{TM}$ is a registered trademark of DWYEROMEGA. NI$^{TM}$ is a registered trademark of National Instruments, Inc. Emerson® is a registered trademark of Emerson, Inc. Cole-Parmer® is a trademark of Cole-Parmer, Inc. Altech® is a trademark of Altech, Inc. Bell & Gossett® is a trademark of Bell & Gosset, Inc.

**Appendix**

**A1. Effect of lengthwise span of cavitation domain**

Naturally, when the occurrence of cavitation in the channel is highly unlikely (i.e. well downstream from the orifice) one does not need to maintain the cavitation terms in the model. The extent to which the accommodation coefficient must be maintained non-zero was thus investigated in this study. Two lengthwise spans from the orifice and up to $5d$ and $10d$ beyond the orifice were considered where the cavitation coefficient was non-zero; the cavitation term was turned off ($C_C$, $C_V$ set to zero) for the remainder of the channel flow beyond the respective cutoff point. The corresponding pressure drop results are illustrated in **Figure** A1(a), where a comparison is made with the scenario where the accommodation coefficient was non-zero throughout the entire domain. As indicated in Figure A1(b), the $10d$ lengthwise span resulted in a pressure drop identical to that of the full-domain case. This finding supports that cavitation effects in the present conditions are localized within a region extending approximately 10 orifice widths downstream of the orifice. Any cavitation considerations beyond this point do not cause any measurable deviation, at least in pressure losses along the channel.

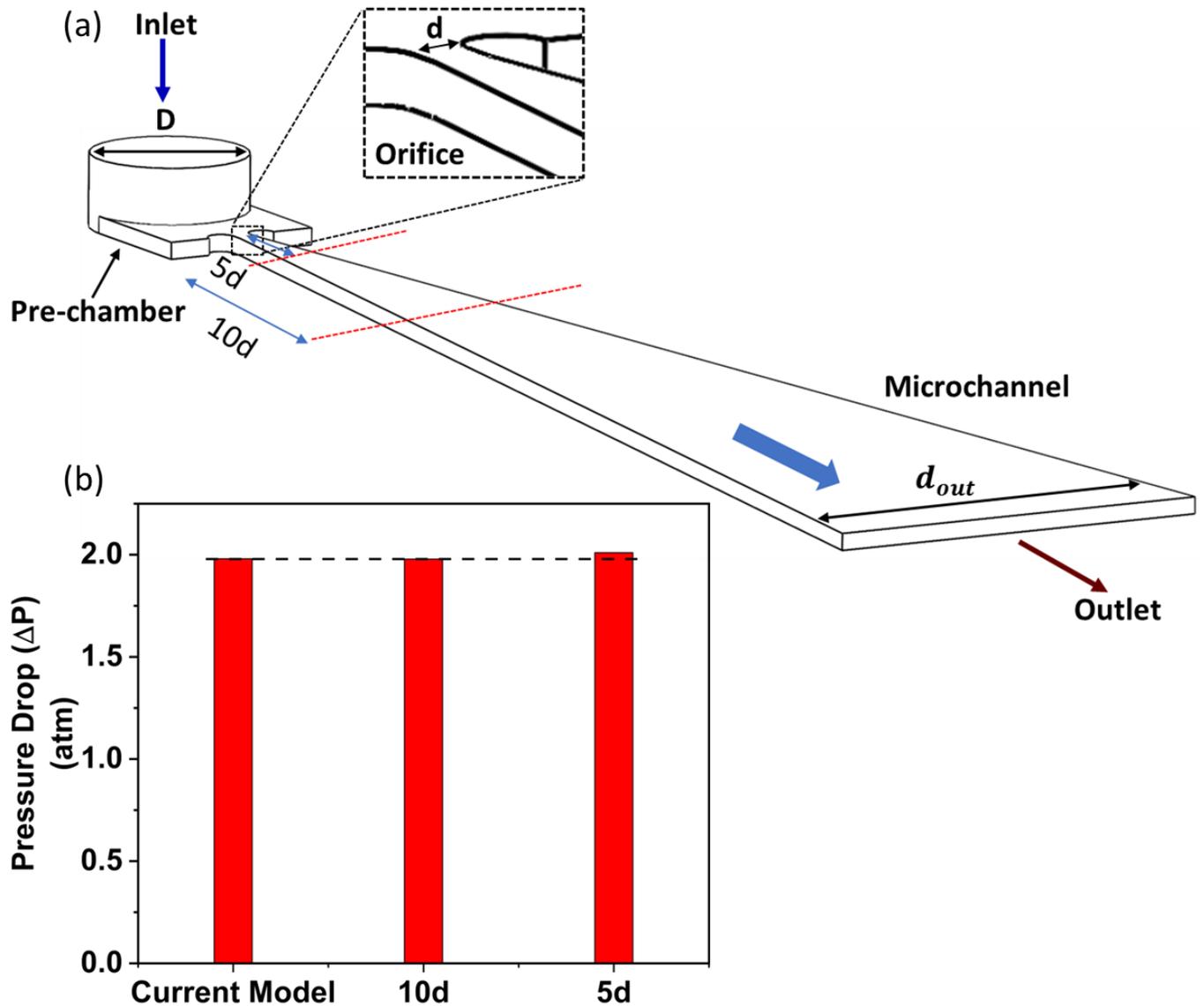

**Figure A1:** (a) Red dashed lines in the cross stream direction mark the ends of the two lengthwise spans (from the orifice to $5d$ or $10d$) where non-zero accommodation coefficients were applied; (b) Comparison of pressure drop for the different spans. The leftmost bar acts as a reference and corresponds to the case where the cavitation coefficient was non-zero for the entire channel length.

**References**

[1] J. Zhang, L. Lei, H. Li, G. Xin, X. Wang, 2022. Experimental and numerical studies of liquid-liquid two-phase flows in microchannel with sudden expansion/contraction cavities. Chem. Eng. J. 433, 133820.

[2] Y. Liao, D. Lucas, 2015. 3D CFD simulation of flashing flows in a converging-diverging nozzle. Nucl. Eng. Des. 292, 149–163.

[3] A. Pal, A. Mukhopadhyay, C.M. Megaridis, 2024. Wick-free vapor chamber featuring laser-textured surfaces: A new paradigm for ultra-thin high-efficiency heat spreaders. Appl. Therm. Eng. 124938.

[4] A. Mukhopadhyay, A. Pal, S. Sarkar, C.M. Megaridis, 2024. Laser-tuned surface wettability modification and incorporation of aluminum nitride (AlN) ceramics in thermal management devices. Adv. Funct. Mater. 2313141.

[5] M. Jafari Gukeh, C. Bao, A. Mukhopadhyay, G. Damoulakis, S.K. Mazumder, C.M. Megaridis, 2023. Air-cooled hybrid vapor chamber for thermal management of power electronics. Appl. Therm. Eng. 224, 120081.

[6] Karmakar S., Pal A., Sarkar S., Mukhopadhyay A., Optimized designs for high-efficiency particle sorting in serpentine microfluidic channels, 2025, Phys. Fluids, 37(4).

[7] M. Hosbach, S. Gitau, T. Sander, U. Leuteritz, M. Pfitzner, 2019. Effect of taper, pressure and temperature on cavitation extent and dynamics in micro-channels. Exp. Therm. Fluid Sci. 108, 25–38.

[8] M. Medrano, J.-P. Franc, M. Mohan, C. Pellone, P.J. Zermatten, F. Ayela, 2010. Cavitation in microchannels. In: 2nd Eur. Conf. Microfluid., pp. 1–10.

[9] W.J. Meng, C. Lei, W.T. Su, B. Li, 2024. Study on microchannel cavitation phenomena based on experiment and simulation. J. Phys. Conf. Ser. 2707, 012126.

[10] M. Yu, X. Peng, X. Meng, J. Jiang, Y. Ma, 2023. Influence of cavitation on the heat transfer of high-speed mechanical seal with textured side wall. Lubricants 11, 378.

[11] M. Hosbach, R. Skoda, T. Sander, U. Leuteritz, M. Pfitzner, 2020. On the temperature influence on cavitation erosion in micro-channels. Exp. Therm. Fluid Sci. 117, 110140.

[12] W. Zhang, S. Li, Y. Ding, M. Duan, F. Liu, 2024. Bubble breakup in microchannels: A review. Chem. Eng. J. 485, 149868.

[13] C.P. Egerer, S. Hickel, S.J. Schmidt, N.A. Adams, 2014. Large-eddy simulation of turbulent cavitating flow in a micro channel. Phys. Fluids 26, 085102.

[14] E. Goncalves, R.F. Patella, 2009. Numerical simulation of cavitating flows with homogeneous models. Comput. Fluids 38, 1682–1696.

[15] Z. Li, Z. Liu, P. Chen, J. Liu, J. Wu, 2022. Numerical comparative study of fuel cavitation in microchannels under different turbulence models. Energies 15, 8265.

[16] M. Hosbach, S. Gitau, T. Sander, U. Leuteritz, M. Pfitzner, 2019. Effect of taper, pressure and temperature on cavitation extent and dynamics in micro-channels. Exp. Therm. Fluid Sci. 108, 25–38.

[17] T. Chen, H. Chen, W. Liang, B. Huang, L. Xiang, 2019. Experimental investigation of liquid nitrogen cavitating flows in converging-diverging nozzle with special emphasis on thermal transition. Int. J. Heat Mass Transf. 132, 618–630.

[18] W. Meng, X. Li, W. Su, 2024. Cavitation characteristics in rectangular flow restriction microchannel seals. Phys. Fluids 36, 123341.

[19] X.Y. Duan, B.H. Huang, Y.X. Zhu, X. Song, C.Y. Zhu, J.C. Chai, L. Gong, 2023. Cavitating flows in microchannel with rough wall using a modified microscale cavitation model. Case Stud. Therm. Eng. 52, 103735.

[20] K. Zhang, J. Yang, X. Huai, 2024. Surface topography controls bubble nucleation at rough water/silicon interfaces for different initial wetting states. Int. J. Heat Mass Transf. 224, 125323.

[21] B. Schneider, A. Koşar, C.J. Kuo, C. Mishra, G.S. Cole, R.P. Scaringe, Y. Peles, 2006. Cavitation enhanced heat transfer in microchannels. J. Heat Transf. 128, 1293–1301.

[22] P. Pipp, M. Hočevar, M. Dular, 2021. Challenges of numerical simulations of cavitation reactors for water treatment – An example of flow simulation inside a cavitating microchannel. Ultrason. Sonochem. 77, 105663.

[23] A. Nayebzadeh, Y. Wang, H. Tabkhi, J.H. Shin, Y. Peles, 2018. Cavitation behind a circular micro pillar. Int. J. Multiph. Flow 98, 67–78.

[24] D. Podbevšek, G. Ledoux, M. Dular, 2022. Investigation of hydrodynamic cavitation induced reactive oxygen species production in microchannels via chemiluminescent luminol oxidation reactions. Water Res. 220, 118628.

[25] A. Pal, A. Mukhopadhyay, C.M. Megaridis, 2025. Wick-free vapor chamber featuring laser-textured surfaces: A new paradigm for ultra-thin high-efficiency heat spreaders. Appl. Therm. Eng. 260, 124938.

[26] A. Mukhopadhyay, A. Pal, C. Bao, M.J. Gukeh, S.K. Mazumder, C.M. Megaridis, 2023. Evaluation of thermal performance of a wick-free vapor chamber in power electronics cooling. In: Proc. ITHERM 2023-May.

[27] M.G. De Giorgi, A. Ficarella, D. Fontanarosa, 2017. Implementation and validation of an extended Schnerr-Sauer cavitation model for non-isothermal flows in OpenFOAM. Energy Procedia 126, 58–65.

[28] M. Nezamirad, A. Yazdi, S. Amirahmadian, N. Sabetpour, A. Hamedi, 2021. Utilization of Schnerr-Sauer cavitation model for simulation of cavitation inception and super cavitation.

[29] X.Y. Duan, B.H. Huang, Y.X. Zhu, X. Song, C.Y. Zhu, J.C. Chai, L. Gong, 2023. Cavitating flows in microchannel with rough wall using a modified microscale cavitation model. Case Stud. Therm. Eng. 52, 103735.

[30] B. Schneider, A. Koşar, Y. Peles, 2007. Hydrodynamic cavitation and boiling in refrigerant (R-123) flow inside microchannels. Int. J. Heat Mass Transf. 50, 2838–2854.

[31] C.P. Egerer, S. Hickel, S.J. Schmidt, N.A. Adams, 2014. Large-eddy simulation of turbulent cavitating flow in a micro channel. Phys. Fluids 26.

[32] G.R. Mondal, M.N. Hossain, A. Kundu, 2024. Numerical analysis of flow condensation inside a horizontal tube: Current status and possibilities. Lect. Notes Mech. Eng. 229–241.

[33] M.N. Hossain, A. Chakravarty, K. Ghosh, 2021. Influence of transient heat input on pool boiling behaviour. In: Proc. ISHMT-ASTFE 2021, IIT Madras, India, pp. 2031–2037.

[34] A. Pal, R. Biswas, S. Sarkar, A. Mukhopadhyay, 2022. Effect of ventilation and climatic conditions on COVID-19 transmission through respiratory droplet transport via both airborne and fomite mode inside an elevator. Phys. Fluids 34, 083319.

[35] A. Pal, R. Biswas, R. Pal, S. Sarkar, A. Mukhopadhyay, 2023. A novel approach to preventing SARS-CoV-2 transmission in classrooms: A numerical study. Phys. Fluids 35, 013308.

[36] H. Jasak, 2009. OpenFOAM: Open source CFD in research and industry. Int. J. Nav. Archit. Ocean Eng. 1, 89–94.

[37] N. Sen, 2021. Transmission and evaporation of cough droplets in an elevator: Numerical simulations of some possible scenarios. Phys. Fluids 33, 033311.

[38] R. Biswas, A. Pal, R. Pal, S. Sarkar, A. Mukhopadhyay, 2022. Risk assessment of COVID infection by respiratory droplets from cough for various ventilation scenarios inside an elevator: An OpenFOAM-based computational fluid dynamics analysis. Phys. Fluids 34, 013318.

[39] F.R. Menter, 1994. Two-equation eddy-viscosity turbulence models for engineering applications. AIAA J. 32, 1598–1605.

[40] G. Schnerr, 2001. Physical and numerical modeling of unsteady cavitation dynamics., ICMF-2001, 4th International Conference on Multiphase Flow New Orleans, USA, May 27 - June 1, 2001, pp. 1–12.

[41] M. Jansson, E. Ghahramani, S. Wadekar, 2018. CFD with opensource software, Implementing a Zwart-Gerber-Belamri cavitation model.

[42] S. Li, J. Yang, Q. Wang, 2017. Large eddy simulation of flow and heat transfer past two side-by-side spheres. Appl. Therm. Eng. 121, 810–819.